\documentclass[twocolumn,showpacs,nofootinbib,tightenlines,nobibnotes, aps, amssymb,amsmath,prb]{revtex4}
\usepackage{graphicx}
\usepackage{epsfig}
\usepackage{amssymb,amsmath,amsfonts,hyperref}
\usepackage{wasysym}
\usepackage{latexsym}
\usepackage{eucal}
\usepackage{verbatim}
\usepackage[normalem]{ulem}

\usepackage{color}

\newcommand{\be}{\begin{eqnarray}}
\newcommand{\ee}{\end{eqnarray}}

\begin{document}
\title{Cluster algorithms for frustrated two dimensional Ising antiferromagnets via dual worm constructions}
\author{Geet Rakala}
\affiliation{\small{Tata Institute of Fundamental Research, 1 Homi Bhabha Road, Mumbai 400005, India}}
\author{Kedar Damle}
\affiliation{\small{Tata Institute of Fundamental Research, 1 Homi Bhabha Road, Mumbai 400005, India}}
\begin{abstract}
We report on the development of two dual worm constructions that lead to cluster algorithms for efficient and ergodic Monte Carlo simulations of frustrated Ising models with arbitrary two-spin interactions that extend up to third-neighbours on the triangular lattice. One of these algorithms generalizes readily to other frustrated systems, such as Ising antiferromagnets on the Kagome lattice with further neighbour couplings. We characterize the performance
of both these algorithms in a challenging regime with power-law correlations at finite wavevector.
\end{abstract}

\vskip2pc

\maketitle
\section{Introduction}

Ising models of ferromagnetism, with ``spins'' $\sigma$ that take on two values $\pm 1$, provide simple examples of systems which undergo a continuous phase transition from a high temperature disordered state to a low temperature state which spontaneously breaks
the global symmetry $\sigma \rightarrow -\sigma$.
The vicinity of this continuous transition poses a challenge to Monte Carlo methods
that rely on local updates. In this critical regime,  the spin correlation length becomes very large, and local updates are unable
to significantly change the state of the system.
In such ferromagnetic Ising models, this ``critical slowing-down'' of local updates
can be combated using the well-known Wolff or Swendsen-Wang cluster algorithms.\cite{Wolff,Swendsen_Wang,Wang_Swendsen}

When the interactions between the spins are antiferromagnetic and the underlying
lattice non-bipartite, the geometry of the lattice causes these antiferromagnetic interactions
to compete with each other. This ``geometric frustration'' is often associated with a macroscopic degeneracy of minimum exchange-energy configurations. This can
lead to interesting liquid-like phases at intermediate temperatures. Subleading further-neighbour interactions can destabilize this liquid to produce complex patterns of low temperature order.
The standard Wolff construction typically fails to yield a satisfactory cluster algorithm in the low
temperature regime of interest in such frustrated systems, a result that
can be understood in terms of the percolation properties of the Wolff clusters.\cite{Leung_Henley} Generalizations\cite{Kandel_Ben-Av_Domany_PRL,Kandel_Domany,Kandel_Ben-Av_Domany_PRB,Coddington_Han,Zhang_Yang}  of the Wolff cluster construction procedure,  which build
clusters by defining a percolation process involving larger units of the lattice (typically,
the elementary plaquettes of the lattice), have also been explored with some success for the fully frustrated Ising model with nearest-neighbour antiferromagnetic exchange
on square and triangular lattices. 
\begin{figure}[t]
  \includegraphics[width=8.4cm]{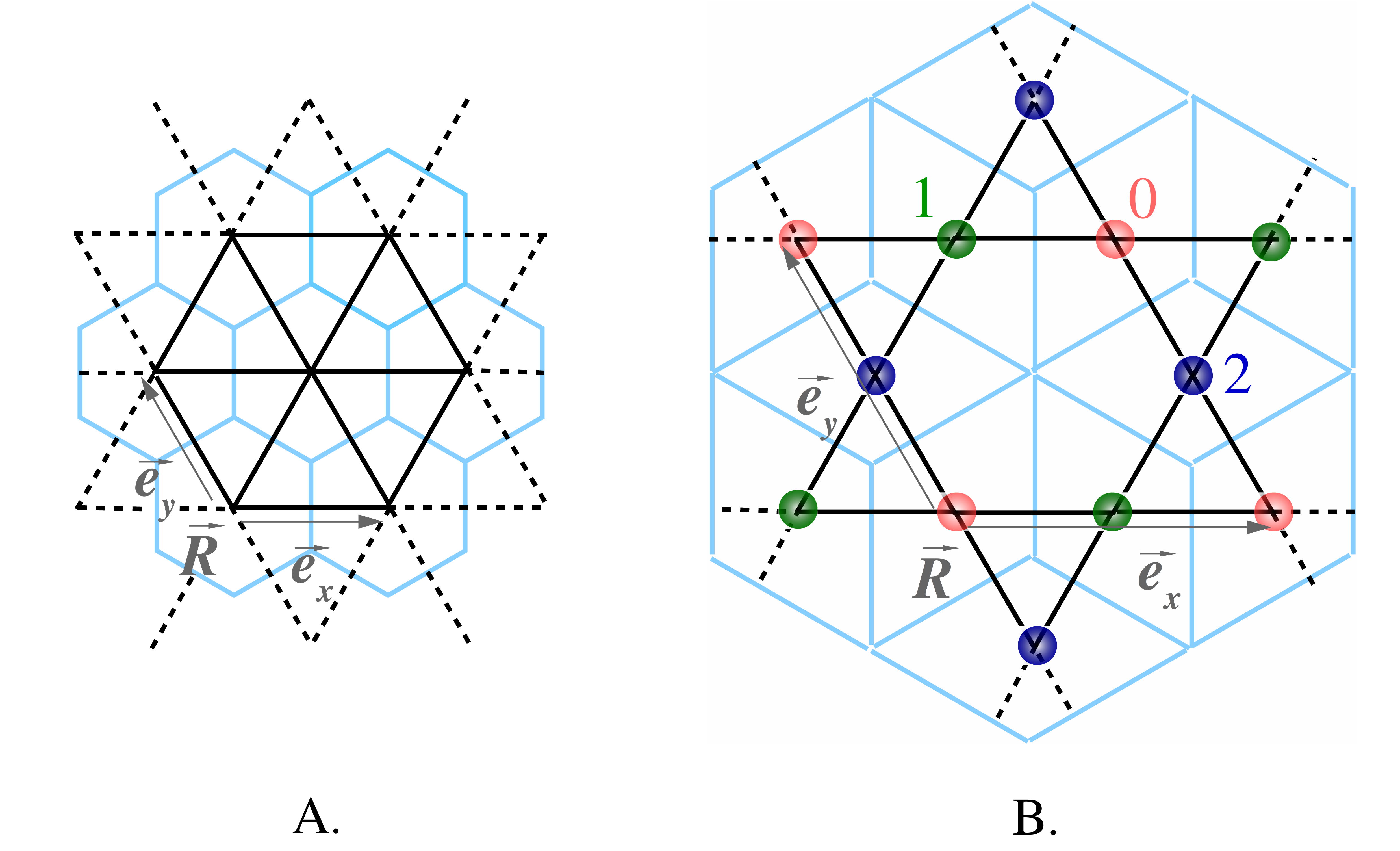}
  \caption{\label{lattice} A) The triangular lattice is shown in black and its dual honeycomb lattice is shown in blue. B) The Kagome lattice is shown in black and its dual dice lattice is shown in blue. Basis sites and Bravais lattice translation vectors are also marked as appropriate.
}
\end{figure}

In ferromagnetic models, a  ``dual worm'' approach has also been used as an alternative to the standard Wolff cluster construction.\cite{Hitchcock_Sorenson_Alet}  This approach
uses a worm construction to effect a nonlocal update in the bond-energies along
a closed loop. When transformed back to spin variables, this dual worm construction
yields a procedure for constructing a spin cluster which can be flipped with probability
one. Recently, such a dual worm algorithm has also been formulated
for some frustrated systems.\cite{Wang_Sterck_Melko}. These generalizations
feature a nonzero rejection rate that is required in order to preserve detailed balance.

Here, we introduce two new cluster algorithms for frustrated triangular lattice Ising models with arbitrary two-spin couplings that extend up to next-next-nearest neighbours.  Both 
algorithms use the dual worm framework of Ref.~\onlinecite{Hitchcock_Sorenson_Alet}. At the heart of these two algorithms are two new worm construction protocols that both preserve detailed balance {\em without any rejection of completed worms both at $T=0$ and $T>0$}. This makes both cluster algorithms very efficient in the entire temperature range of interest in such frustrated magnets.

In the limiting case of $J_1 \rightarrow \infty$, or equivalently, in
the $T\rightarrow 0$ limit with $J_1=1$ and $J_{2/3} = c_{2/3} T$ (with constant $c_2$ and $c_3$), one of these algorithms, which we dub the ``DEP'' algorithm (since it
involves deposition, evaporation, and pivoting of dual dimer variables), reduces
to a previously-used\cite{Sen_Damle_Vishwanath,Sen_thesis,Patil_Dasgupta_Damle,Mila_etal}  honeycomb lattice implementation of the well-known worm algorithm for interacting dimer models\cite{Hitchcock_Sorenson_Alet}. In the same limit, our other algorithm, which we dub
the ``myopic'' algorithm since it ignores the environment variables at alternate
steps of the worm construction, reduces to the honeycomb lattice implementation
of an approach developed
previously as an alternative\cite{KDamle_linkcurrents} to the standard procedure\cite{Alet_linkcurrents} for constructing worm updates for generalized link-current models.

This myopic algorithm generalizes readily to other frustrated systems, including Kagome lattice Ising antiferromagnets with two-spin couplings again extending up to next-next-nearest neighbours. In the limiting case of $J_1 \rightarrow \infty$ on the Kagome lattice (equivalently, in the $T\rightarrow 0$ limit with $J_1=1$ and $J_{2/3} = c_{2/3} T$ with constant $c_2$ and $c_3$), this myopic algorithm reduces to an approach used in previous work to simulate an interacting dimer model on the dice lattice.\cite{Sen_Wang_Damle_Moessner,Sen_thesis}

In this paper, we present a detailed characterization of both these cluster algorithms in 
the low temperature power-law ordered intermediate state associated with the two-step thermal melting of three-sublattice
order in such triangular lattice Ising models. Similar results are also obtained
in the Kagome lattice case with the myopic algorithm. We demonstrate that both
cluster algorithms have a significantly smaller dynamical exponent $z$ compared to that of the standard Metropolis algorithm. Interestingly, we also find that the dependence of the dynamical exponent on the equilibrium anomalous dimension $\eta$ is quasi-universal in a sense that we attempt to make precise in this work. The rest of this paper is organized as follows: In Section~\ref{Models}, we introduce the Ising models of interest to us and summarize the basic physics of three-sublattice ordering
in such triangular lattice antiferomagnets. In Section~\ref{Algorithms}, we provide a  detailed description of the two cluster algorithms developed here, and present benchmarks establishing the correctness of the procedures used. In Section.~\ref{Performance}, we provide a characterization of the performance of our cluster algorithms, and compare this performance to that of the standard Metropolis algorithm. We conclude in Section.~\ref{Outlook} with a brief discussion.

\section{Models}
\label{Models}
Ising antiferromagnets\cite{Wannier,Stephenson,Kano_Naya}  on frustrated lattices such as the triangular and the Kagome
lattice provide paradigmatic examples of the effects of geometric frustration
in low dimensional magnets. In such systems, the behaviour at
low temperature is governed by the  interplay between the macroscopic
degeneracy of configurations that minimize the nearest-neighbour antiferromagnetic
exchange energy, and subleading energetic preferences imposed by weaker further-neighbour interactions.
The classical Hamiltonian for these model systems on the triangular and Kagome lattices may be written as
\begin{eqnarray}
H&=&J_1\sum_{\langle RR' \rangle} \sigma_R \sigma_{R'} + J_2 \sum_{\langle \langle RR' \rangle \rangle} \sigma_R \sigma_{R'} + J_3  \sum_{\langle \langle \langle RR' \rangle \rangle \rangle} \sigma_R \sigma_{R'} \;  , \nonumber \\
&&
\end{eqnarray}
where $\langle R R' \rangle$, $\langle \langle R R' \rangle \rangle$, and $\langle \langle \langle R R' \rangle \rangle \rangle$ denote nearest neighbour, next-nearest neighbour,
and next-next-nearest neighbour links of the lattice in question, and $\sigma_R = \pm 1$
are the Ising spins at sites $R$ of this lattice. In our convention, $J_{1/2/3} >0$ corresponds
to an antiferromagnetic coupling, while $J_{1/2/3} < 0$ corresponds to  a ferromagnetic
coupling. In the rest of this paper, $J_1$ is assumed positive and equal to $1$. 

When $J_2=J_3=0$, the nearest-neighbour model does not order on either lattice even in the zero temperature limit, providing an example of a classical spin liquid state. On the triangular lattice, the
correlation length grows exponentially with
decreasing temperature, reflecting the fact that the $T=0$ spin-liquid,
which involves an average over the ensemble of
minimum nearest-neighbour exchange energy states, is characterized
by power-law spin correlations at the three-sublattice wavevector ${\mathbf Q}$.\cite{Wannier,Stephenson}
On the Kagome lattice, $H$ with $J_2=J_3=0$ remains a short-range correlated
spin liquid all the way down to zero temperature.\cite{Kano_Naya} 

When $J_2$ is negative, such magnets tend to develop three-sublattice spin
order at low temperature. In this ordered state, the spins freeze in
a pattern that is commensurate with the three-sublattice decomposition
of the underlying triangular Bravais lattice. This parameter regime
has attracted some interest earlier in the context of spatial ordering of
monolayer adsorbate films on substrates with triangular symmetry,\cite{Bretz_etal,Bretz,Horn_Birgeneau_Heiney_Hammonds,Vilches,Suter_Colella_Gangwar,Feng_Chan,Wiechert}  and in
the context of ``artificial spin-ice'', {\em i.e.} honeycomb networks of micromagnetic
wires which can be modeled in terms of the Kagome lattice Ising antiferromagnet
with further neighbour couplings.\cite{Tanaka_etal,Qi_etal,Ladak_etal1,Ladak_etal2}
In both these examples, the three-sublattice order is of the {\em ferrimagnetic}
type, {\em i.e.}, it is accompanied by a small net moment.

As a computationally challenging regime in which to test our algorithms, we focus here on this ferrimagnetc three-sublattice ordered state
that is stabilized at low enough temperature by a nonzero ferromagnetic $J_2$ ($J_2 <0$) on both lattices\cite{Landau,Nienhuis_Hilhorst_Blote,Wolf_Schotte,Takagi_Mekata,Wills_Ballou_Lacroix} (for a simple caricature
of this state on both lattices, see Fig.~1 of Ref.~\onlinecite{KDamle_PRL}). On both
lattices, there is a large range of parameters for which this three-sublattice order melts in a two-step manner on heating,\cite{Landau,Nienhuis_Hilhorst_Blote,Wolf_Schotte,Takagi_Mekata,Wills_Ballou_Lacroix} via an intermediate phase with power-law three-sublattice order corresponding to a temperature dependent power-law exponent $\eta \in (\frac{1}{9}, \frac{1}{4})$. In our work here, we test our algorithms in this extended critical region and extract
the values of the dynamical critical exponents that characterize our algorithms. However, we re-emphasize that the algorithms developed here have much wider applicability,
and work well for arbitrary values of $J_2$ and $J_3$ on both lattices when the nearest-neighbour interaction $J_1$ is positive.

\section{Algorithms}
\label{Algorithms}

\begin{figure}[t]
  \includegraphics[width=8.4cm]{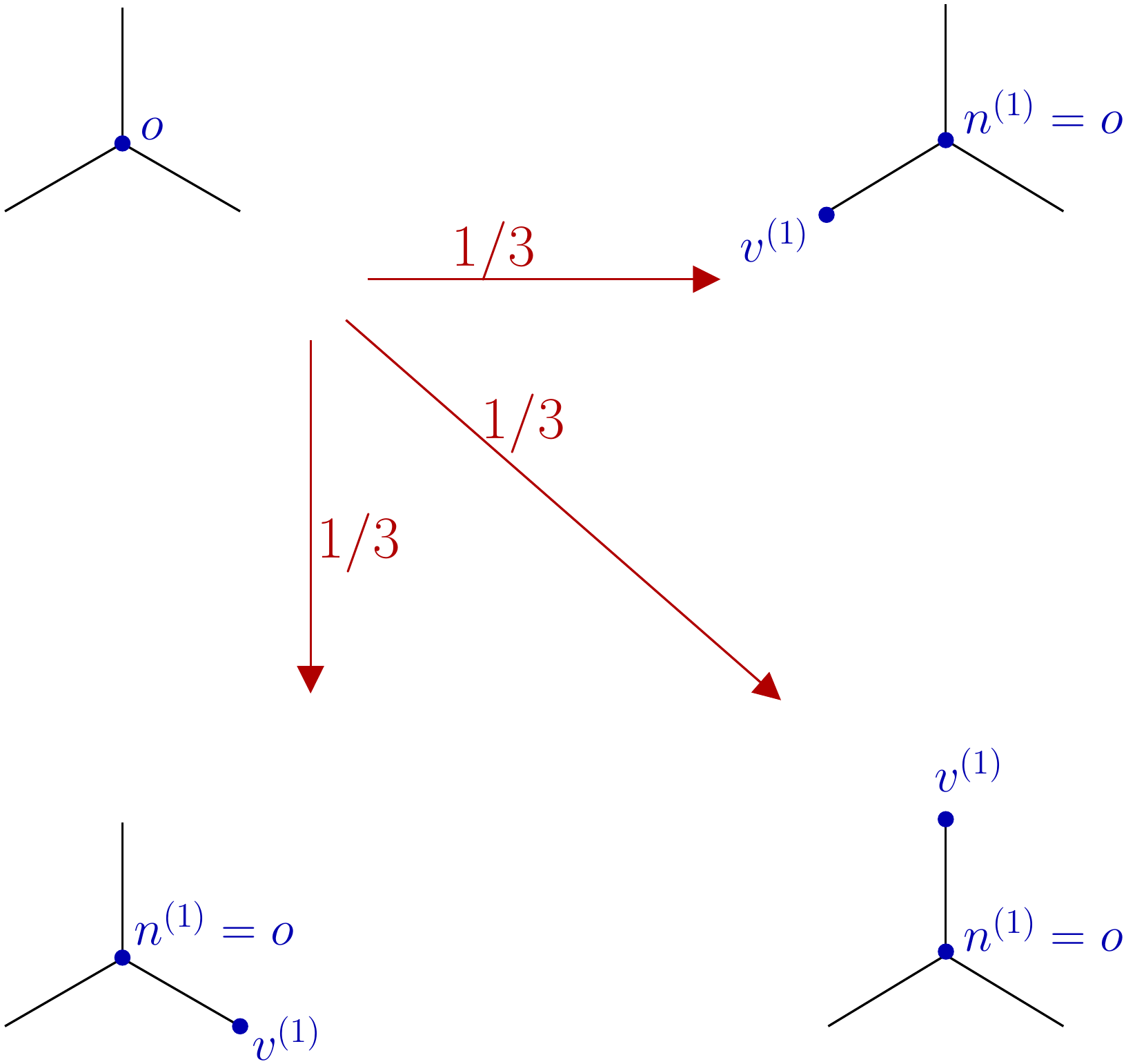}
  \caption{\label{start_myopic_triangular} The first step of the myopic worm construction on the dual honeycomb lattice: A start site $o$ is chosen randomly. The tail of the worm remains static
at this start site until the worm construction is complete. In this first step, the head of the worm moves to one of the three neighbours of the start site with probability $1/3$, regardless of the local dimer configuration. The neighbour thus reached is our first vertex site $v^{(1)}$. Viewed from the point of view of this vertex, the start site $o$ is the first entry site $n^{(1)}$.
}
\end{figure}

\subsection{Dual Representation and dual worm updates}
We begin with a summary of the dual representation of
the frustrated Ising antiferromagnet on the triangular lattice: One represents every configuration of the triangular lattice Ising model in terms of configurations of dimers on links of the dual honeycomb lattice, with
a dimer placed on every dual link that intersects a {\em frustrated} nearest neighbour
bond of the triangular lattice (Fig.~\ref{lattice}). For our purposes here, a frustrated bond of the triangular
lattice is one that connects a pair of spins pointing in the same direction. When
$J_1 > 0$ (corresponding to the interesting case of frustrated antiferromagnetism), this
implies that every {\em minimally frustrated} spin configuration, which minimizes the
nearest-neighbour exchange energy by ensuring that every triangle of
the triangular lattice has exactly one frustrated bond, corresponds to a defect-free dimer cover of the dual honeycomb lattice, in which there is exactly one dimer touching each dual site of the honeycomb lattice.

At non-zero temperature,
more general configurations also contribute to the partition sum.  These have a nonzero density of {\em defective triangles}, {\em i.e.} triangles in which all three spins are pointing
in the same direction. In dual language, these correspond to honeycomb lattice sites
with three dimers touching the site. Thus, in dual language, the configuration space at nonzero temperature is that of a generalized honeycomb lattice dimer model, with either one or three dimers touching each dual site. This dimer model inherits boundary conditions
from the original spin model: We choose to work with $L_x \times L_y$ samples
with periodic boundary conditions on the Ising spins along two principal directions $\hat{x}$ and $\hat{y}$ of the triangular lattice. This translates to global constraints on the dual description which are spelled out in detail when we describe our algorithm.

All of this generalizes readily to the Kagome lattice antiferromagnet. The idea
is again to work with the dual representation in terms of a generalized dimer model
on the dual lattice. The dual lattice is now the dice lattice, which
is a bipartite lattice with one sublattice of three-coordinated
sites and a second sublattice of six-coordinated sites (Fig.~\ref{lattice}). Every spin configuration
on the Kagome lattice corresponds to a dimer configuration on the dice lattice, with either one or three dimers touching the three-coordinated sites, and an even number of dimers
touching the six-coordinated sites. As before, a frustrated
bond is one that connects a pair of nearest neighbour spins pointing in the same direction, and is
represented by a dimer on the dual link that is perpendicular to this bond. Minimally
frustrated spin configurations, that minimize the nearest-neighbour exchange energy,
now correspond to dimer configurations with exactly one dimer touching each
three-coordinated dice lattice site. Periodic boundary conditions of the $L_x \times L_y$ spin system again translate to global constraints (see below). 
\begin{figure}[t]
  \includegraphics[width=8.4cm]{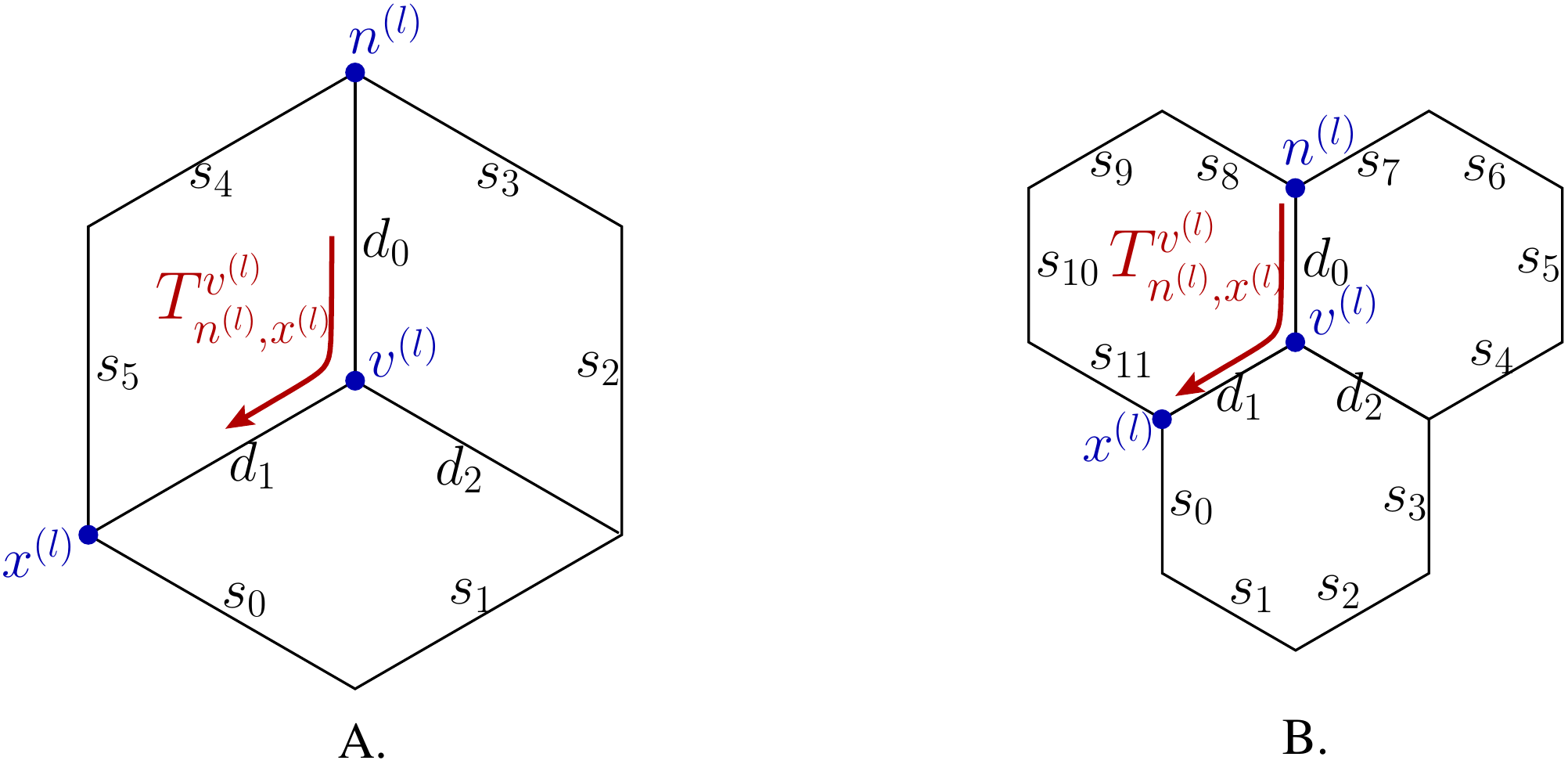}
  \caption{\label{myopic_3by3} The probabilistic step of the myopic algorithm: After arriving at the vertex site $v^{(l)}$ from the entry site $n^{(l)}$ we choose to exit via  $x^{(l)}$ (which is one of the neighbours of $v^{(l)}$) with probability given by the probability table $T$. A) When the spin interactions extend upto next-next-nearest neighbours on the Kagome lattice, knowledge of the local dimer configuration consisting of dimer states from $s_0$ to $s_5$ and $d_0$, $d_1$ and $d_2$ suffices to compute entries of the table $T$.  B) When the spin interactions extend upto next-next-nearest neighbours on the triangular lattice, knowledge of the local dimer configuration consisting of dimer states from $s_0$ to $s_{11}$ and $d_0$, $d_1$ and $d_2$ suffices to to compute entries of $T$.
}
\end{figure}
\begin{figure}[t]
  \includegraphics[width=8.4cm]{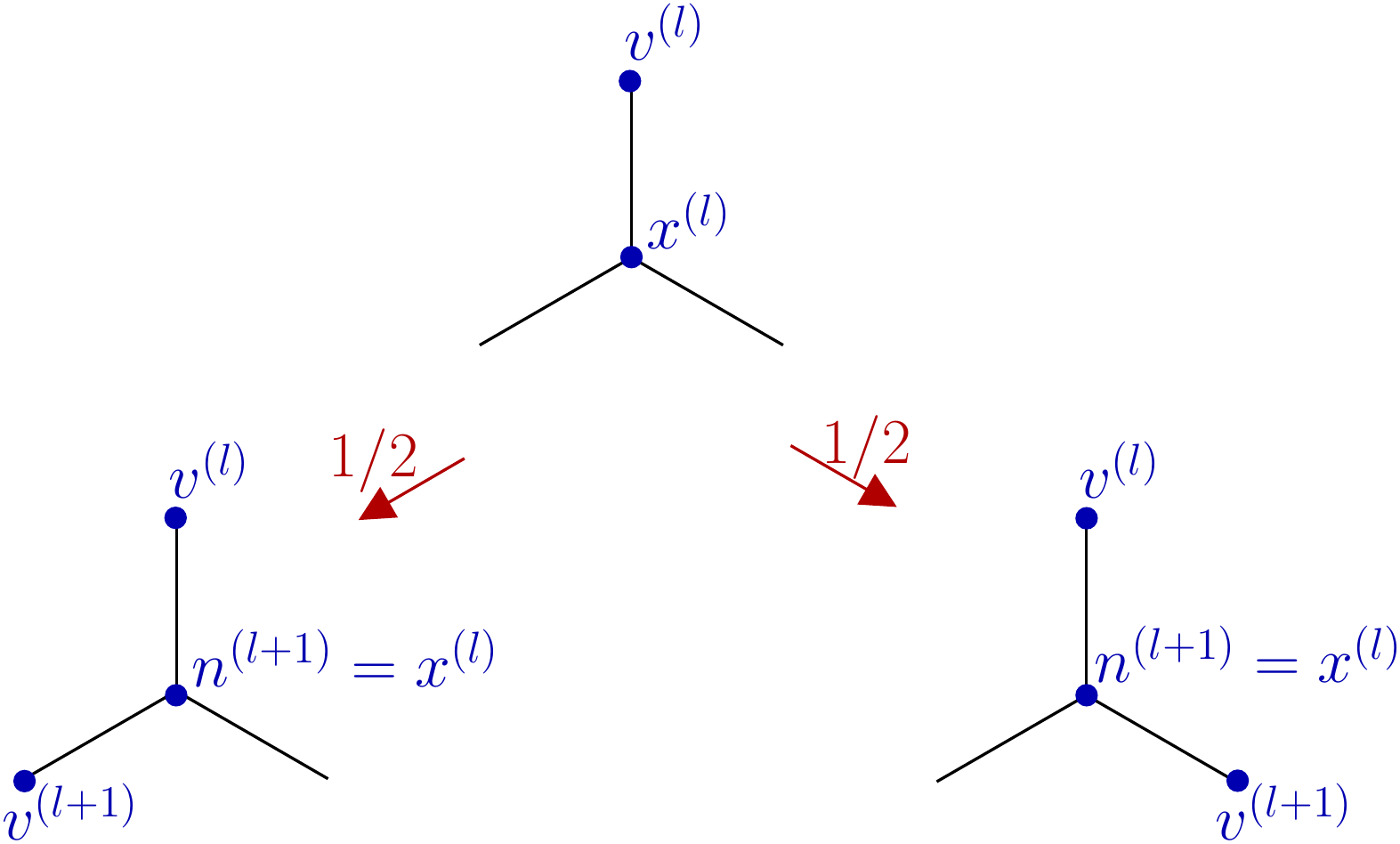}
  \caption{\label{myopic_triangular} The {\em myopic} step of the myopic worm algorithm on the dual honeycomb lattice: After arriving at an exit site $x^{(l)}$ from a vertex site $v^{(l)}$, the next vertex site $v^{(l+1)}$ is chosen to be one of the two {\em other} neighbours of $x^{(l)}$ with probability $1/2$. Viewed from this new vertex site $v^{(l+1)}$, $x^{(l)}$ becomes
the entry site $n^{(l+1)}$.  
}
\end{figure}

The dual worm approach,\cite{Hitchcock_Sorenson_Alet} on which both cluster algorithms are based, is rather simple to explain in general terms: One first maps
the spin configuration of the system to the corresponding dual configuration of dimers.
Each dimer configuration is thus assigned a Boltzmann weight of the ``parent''
spin configuration from which it was obtained.
Next, one updates the dual dimer configuration in a way that preserves detailed balance.
In this way, one obtains a new dimer configuration, which is then checked to see if it satisfies certain global winding number constraints (spelled out in detail below) that must be obeyed by any dimer configuration that is dual to a spin configuration with periodic boundary conditions. If the global constraints are satisfied, one maps
the new dimer configuration back to spin variables, to obtain an updated spin configuration
which can differ from the original spin configuration by large nonlocal changes. Since the
procedure explicitly satisfies detailed balance, one obtains in this way a valid algorithm for the spin model being studied.

For the triangular lattice Ising antiferromagnet, we have developed two strategies for constructing rejection-free updates of the generalized dimer model on the dual honeycomb lattice. As mentioned in the Introduction, one of these
generalizes readily to the generalized dice lattice dimer model which is dual to the frustrated Kagome lattice Ising model, while the other is specific to the generalized honeycomb
lattice dimer model.

The strategy that generalizes readily to the dice lattice case is one in which we deliberately {\em do not}
keep track of the local dimer configuration of the dual lattice at alternate steps of the worm construction in order to ensure that detailed balance can be satisfied without
any final rejection step. This is similar to the myopic worm construction developed
earlier\cite{KDamle_linkcurrents} for a general class of link-current models\cite{Alet_linkcurrents}, and used successfully on the dice lattice in earlier work on an interacting dimer model for the low temperature properties of certain high-spin Kagome lattice antiferromagnets with strong easy-axis anisotropy.\cite{Sen_Wang_Damle_Moessner}
Since this strategy involves being
deliberately short-sighted at alternate steps of the construction, we dub this the ``myopic'' worm algorithm. Below we begin with
a detailed description of how this works for the generalized dimer models
on the honeycomb and dice lattices which are dual to the physics of the frustrated
Ising models on the triangular and Kagome lattices.
 \begin{figure}[t]
  \includegraphics[width=8.4cm]{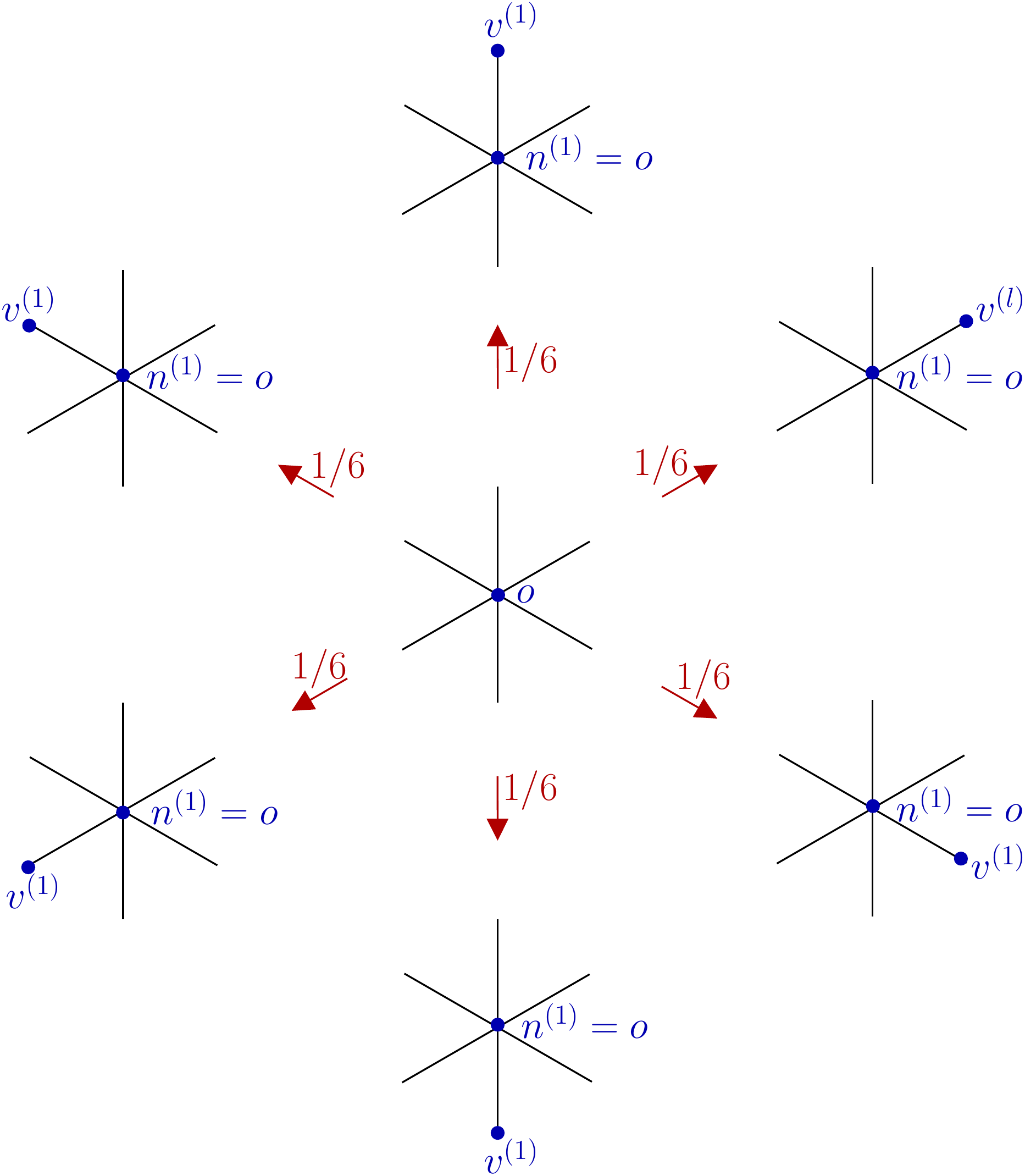}
  \caption{\label{start_myopic_kagome} 
The first step of the myopic worm construction on the dual dice lattice: A start site $o$ is chosen randomly from one of the six-coordinated sites on the dice lattice. The tail of the worm remains static
at this start site until the worm construction is complete. In this first step, the head of the worm moves to one of the six neighbours of the start site with probability $1/6$, regardless of the local dimer configuration. The neighbour thus reached is our first vertex site $v^{(1)}$. Viewed from the point of view of this vertex, the start site $o$ is the first entry site $n^{(1)}$.
}
\end{figure}
\begin{figure}[t]
  \includegraphics[width=8.4cm]{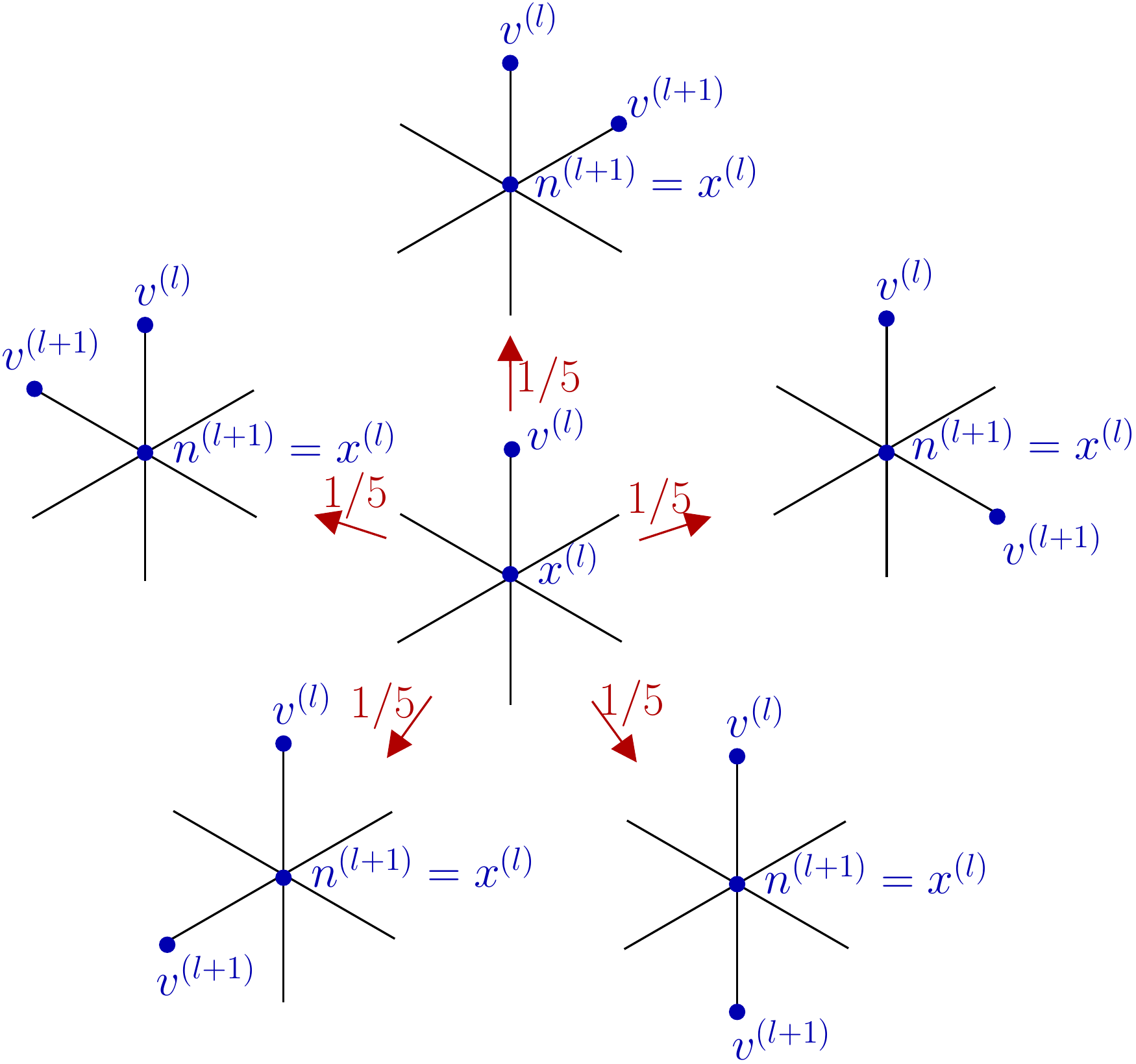}
  \caption{\label{myopic_kagome} The {\em myopic} step of the myopic worm algorithm on the dual dice lattice: After arriving at an exit site $x^{(l)}$ (which, by construction, is always a six-coordinated site) from a vertex site $v^{(l)}$, the next vertex site $v^{(l+1)}$ is chosen to be one of the five {\em other} neighbours of $x^{(l)}$ with probability $1/5$. Viewed from this new vertex site $v^{(l+1)}$, $x^{(l)}$ becomes
the entry site $n^{(l+1)}$.  
}
\end{figure}

\begin{figure}[t]
  \includegraphics[width=8.4cm]{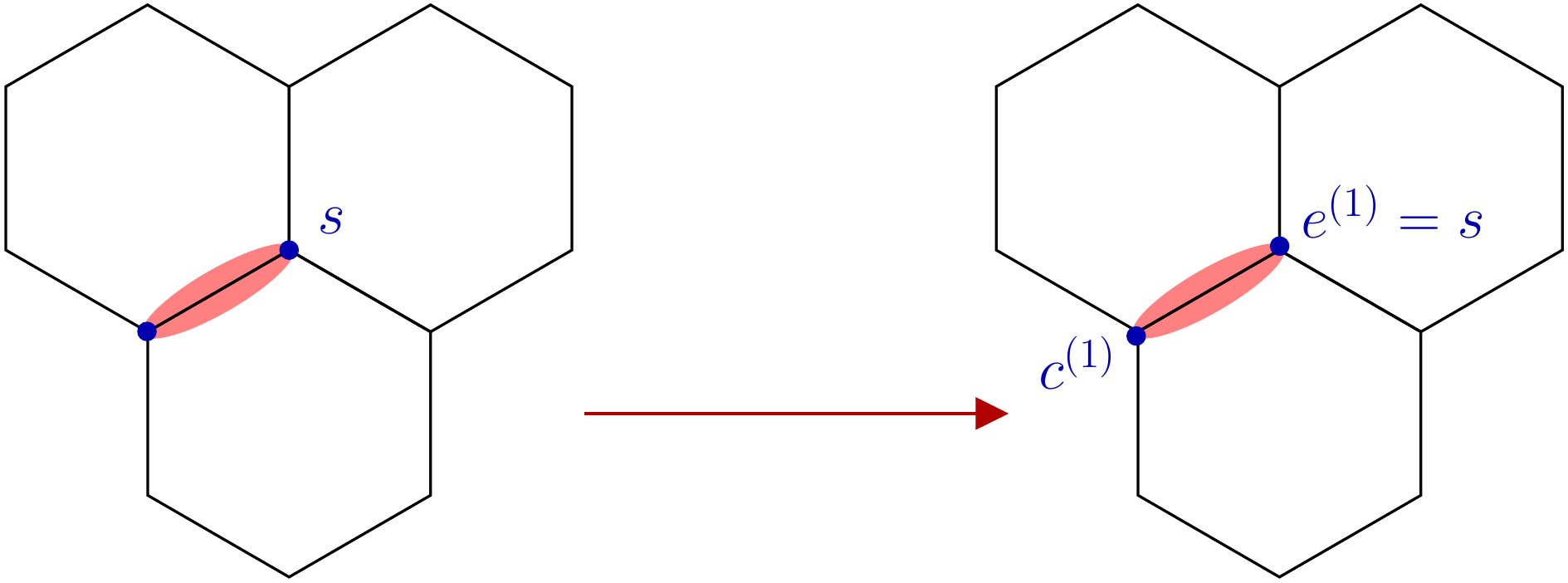}
  \caption{\label{start_branch1} If the randomly chosen start site $s$ is touched by only one dimer, we move along that dimer to reach a new pivot site $c^{(1)}$ in the first step of the DEP worm construction. The start site $s$ becomes our first entry site $e^{(1)}$. 
}
\end{figure}
\begin{figure}[t]
  \includegraphics[width=8.4cm]{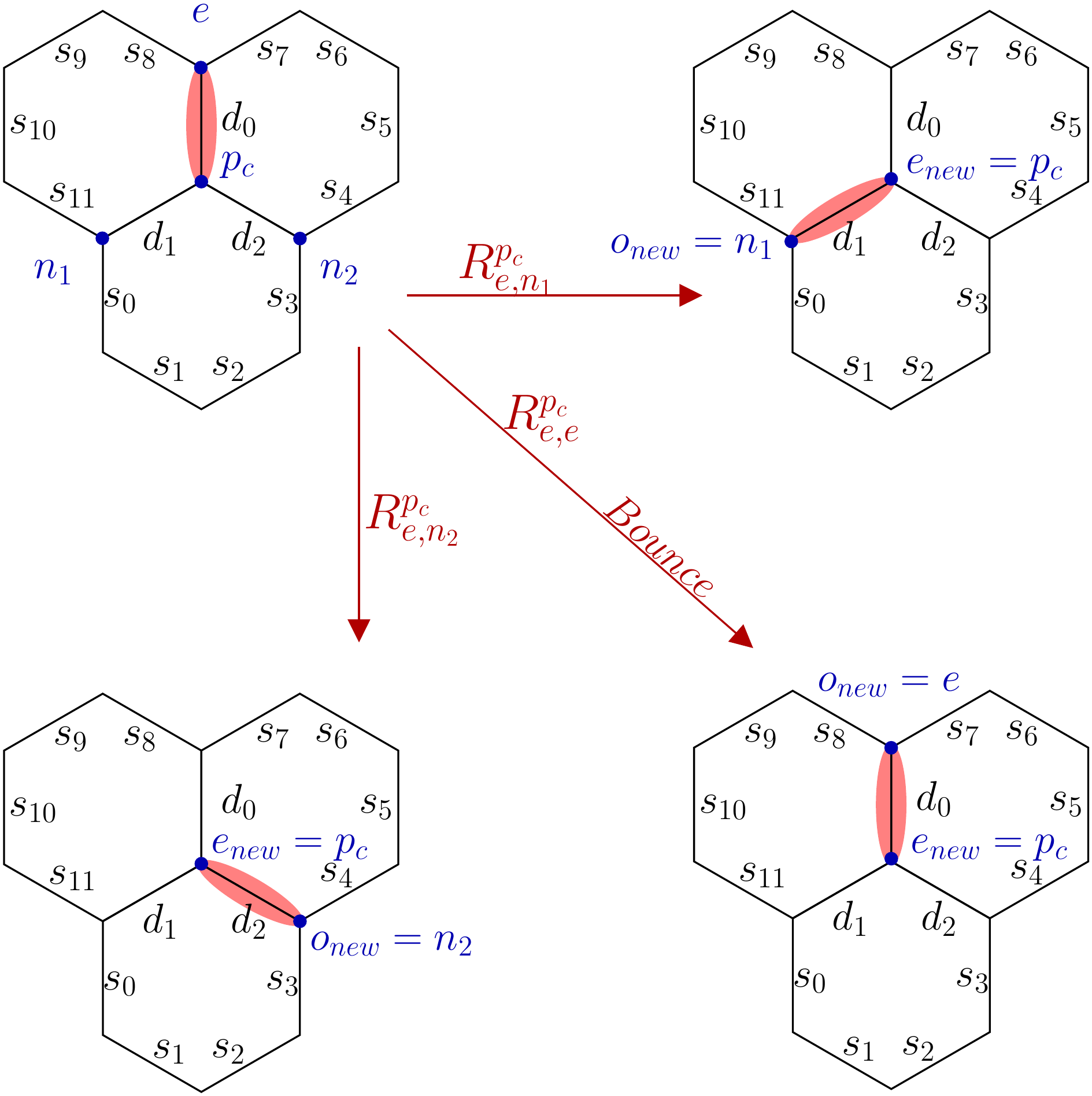}
  \caption{\label{loop} The pivot step of the DEP worm construction when one arrives at a central pivot site $p_c$ from an entry site $e$ and there is only one dimer touching $p_c$. We pivot the dimer from the link $\langle p_c e\rangle$ to the link $\langle p_c o_{new}\rangle$ with probabilities determined by the corresponding elements of the probability table $R$. $o_{new}$, which is the new overlap site, can either one of the two neighbours $n_1$ and $n_2$ of the central pivot site $p_c$ or the entry site $e$ from which we came to $p_c$. The central pivot site $p_c$ now becomes the new entry site $e_{new}$ from which this new overlap site $o_{new}$has been reached, and the next step is an overlap step. On the dual honeycomb lattice, knowledge of the local dimer configuration consisting of dimer states from $s_0$ to $s_{11}$ and $d_0$, $d_1$ and $d_2$ suffices to calculate $R$ when the interactions extend up to next-next-nearest neighbours on the triangular lattice.
}
\end{figure}

\subsection{Myopic worm algorithm}

On the honeycomb lattice this myopic worm algorithm consists of the following steps: We begin by choosing a random ``start site'' $o$ on the honeycomb lattice.  Regardless of the local dimer configuration in the vicinity of this site, we move
from the start site to one of the three neighbouring sites, with probability $1/3$ each (Fig.~\ref{start_myopic_triangular}).
The neighbouring site reached in this way is our first ``vertex site'' $v^{(1)}$. In our
terminology, we have ``entered'' this
vertex site from the start site $o$. Therefore, the start site is the ``entry site'' $n^{(1)}$ for
this vertex. Next, we choose one of the neighbours of $v^{(1)}$ as the ``exit site'' $x^{(1)}$, via which we can exit this vertex.  When we
arrive at vertex site $v^{(1)}$ from entry site $n^{(1)}$, and leave this vertex site via exit site $x^{(1)}$,
we flip the dimer state of the dual links $\langle n^{(1)}v^{(1)} \rangle $ and $\langle v^{(1)}x^{(1)} \rangle$.
The choice of exit site $x^{(1)}$ via which we exit from a vertex site $v^{(1)}$, given that we
arrived at vertex site $v^{(1)}$ from a particular entry site $n^{(1)}$, is probabilistic (Fig.~\ref{myopic_3by3}), with
probabilities specified in a probability table $T$ whose structure we now discuss. 
\begin{figure}[t]
  \includegraphics[width=8.4cm]{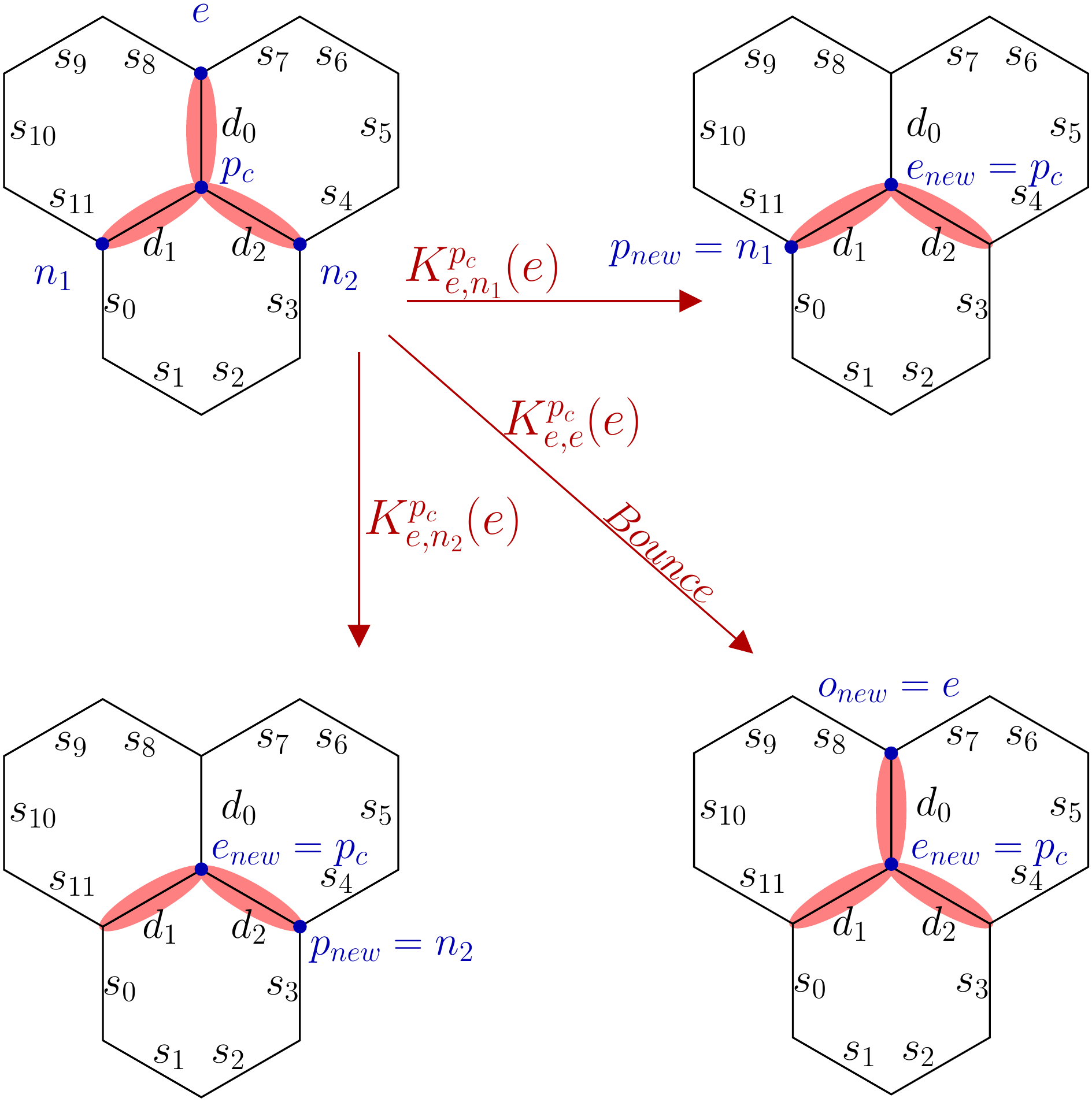}
  \caption{\label{kill} The pivot step of the DEP worm construction when one arrives at a central pivot site $p_c$ from an entry site $e$ and there are three dimers touching $p_c$: At this point,
one has three options, with probabilities determined by corresponding entries of the probability table $K$: We can choose to exit to one of the two neighbours $n_1$ or $n_2$, or bounce back to the entry site $e$. If we choose to exit through either $n_1$ or $n_2$, we move along the dimer connecting the central pivot site $p_c$ to this chosen exit which becomes our new pivot site $p_{new}$, and delete the dimer on the link $\langle p_c e \rangle$. The third option is to move along the dimer connecting the central pivot site $p_c$ back to the entry site $e$, and $e$ then becomes our new overlap site $o_{new}$. On the dual honeycomb lattice, knowledge of the local dimer configuration consisting of dimer states from $s_0$ to $s_{11}$ and $d_0$, $d_1$ and $d_2$ suffices to determine $K$ when the interactions extend upto next-next-nearest neighbours on the triangular lattice. 
}
\end{figure}
\begin{figure}[t]
  \includegraphics[width=8.4cm]{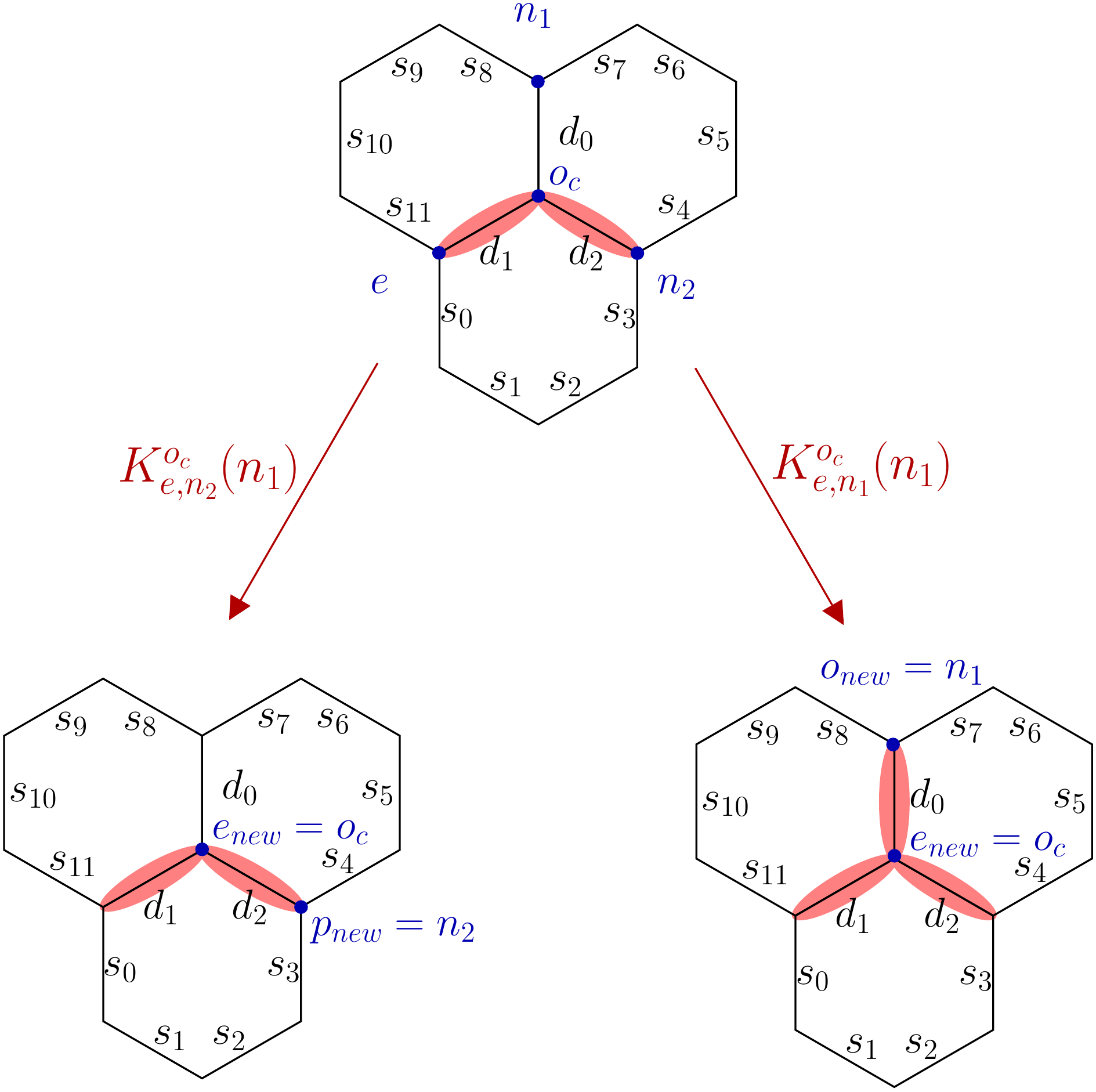}
  \caption{\label{make} The overlap step of the DEP worm construction: One arrives at a central overlap site $o_c$ from an entry site $e$. At this stage, one uses the probability table $K$ to choose one out of two options: If we choose to exit along the empty link (in this case $\langle o_c n_1 \rangle$), we {\em deposit} a dimer on the empty link, and move along it making $n_1$ our new overlap site $o_{new}$. The central overlap site $o_c$ now becomes the new entry site $e_{new}$ from which we enter the new overlap site $o_{new}$. On the other hand, we may choose to exit along the link $\langle o_c n_2 \rangle$ to reach our new pivot site $p_{new} = n_2$. The central overlap site $o_c$ now becomes the new entry site $e_{new}$, from which we enter the new pivot site $p_{new}$. On the dual honeycomb lattice, knowledge of the local dimer configuration consisting of dimer states from $s_0$ to $s_{11}$ and $d_0$, $d_1$ and $d_2$ suffices to determine the probability table $K$, when the interactions extend upto next-next-nearest neighbours on the triangular lattice. 
}
\end{figure}

For any
vertex site $v$ encountered in our process, these probabilities are given by a probability table $T^{v}_{n   x}$, where $n$ is the entry site from which we have entered
the vertex and $x$ is the exit site we wish to leave from. 
Entries in this probability table are constrained by the requirement of local detailed
balance. In order to state these constraints on $T^{v}$ in a way that makes
subsequent analysis easy, we rewrite this table as a three-by-three matrix $M^{v}_{ij}$ ($i, j = 1,2,3$) by choosing a standard
convention to label the three neighbours of $v$ by integers running from one to three.
Thus, if $n$ is the $i^{\mathrm{th}}$ neighbour of $v$ and $x$ is the $j^{\mathrm{th}}$ neighbour
of $v$ according to this convention, we write $T^{v}_{n   x} = M^{v}_{i j}$.

We denote by $w^{v}_n$ the Boltzmann weight of the dual dimer configuration
before we flip the dimer states of dual links $\langle nv \rangle $ and $\langle vx \rangle$. In the same way, $w^{v}_x$, for each choice of $x$, denotes the corresponding Boltzmann weight after these flips are implemented. As is usual for all worm algorithms, these weights for the intermediate configurations encountered during this myopic construction 
are obtained from the Boltzmann weight of the generalized dimer model with
the proviso that the ``infinite energy cost'' of violating the generalized dimer constraints at
the start site and current site (``head'' and ``tail'' of the worm in worm algorithm parlance)
are ignored when keeping track of the weights of these intermediate configurations.

We choose the $T$ matrices to satisfy a local detailed balance condition that depends
on these weights
\begin{eqnarray}
w^{v}_n T^{v}_{n   x} &=& w^{v}_x T^{v}_{x   n} \; .
\end{eqnarray}
Rewriting $w^{v}_n \equiv W^{v}_{i}$ if $n$ is the $i^{\mathrm{th}}$ neighbour
of $v$, and $w^{v}_x \equiv W^{v}_j$ if $x$ is the $j^{\mathrm{th}}$ neighbour
of $v$, we can write these detailed balance conditions
in terms of the matrix $M^{v}_{ij}$ and the weights $W^{v}_{i}$ (with $i,j = 1,2,3$) as
\begin{eqnarray}
W^{v}_i M^{v}_{ij} = W^{v}_j M^{v}_{ji} \; ,
\end{eqnarray}

As is usual in the analysis of such detailed balance constraints\cite{Syljuasen_Sandvik,Syljuasen}, we define the three-by-three matrix $A^{v}_{i j} = W^{v}_i M^{v}_{ij}$ and note that the detailed
balance condition is now simply the statement that $A^v$ is a symmetric matrix which
satisfies the three constraints
\begin{eqnarray}
\sum_j A^{v}_{ij} &=& W^{v}_i \; {\mathrm for} \; i=1,2,3 \nonumber \\
&&
\end{eqnarray}

\begin{figure}[t]
  \includegraphics[width=8.4cm]{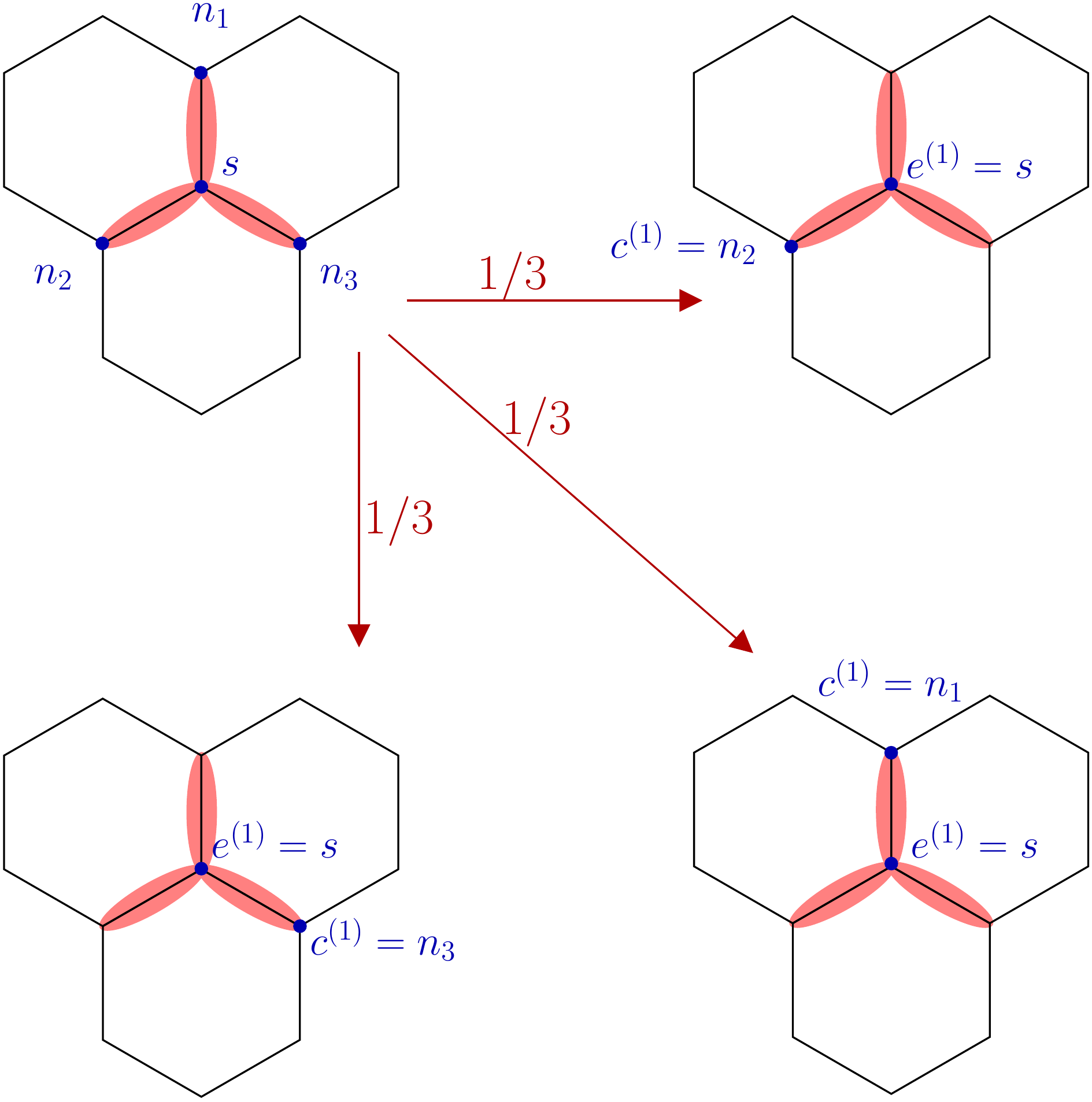}
  \caption{\label{start_branch2} The first step of the DEP worm construction when the randomly chosen start site $s$ is touched by three dimers: We move along any one of the three dimers with probability $1/3$, to reach a new pivot site $c^{(1)}$. The start site $s$ becomes our first entry site $e^{(1)}$.
}
\end{figure}
\begin{figure}[t]
  \includegraphics[width=8.4cm]{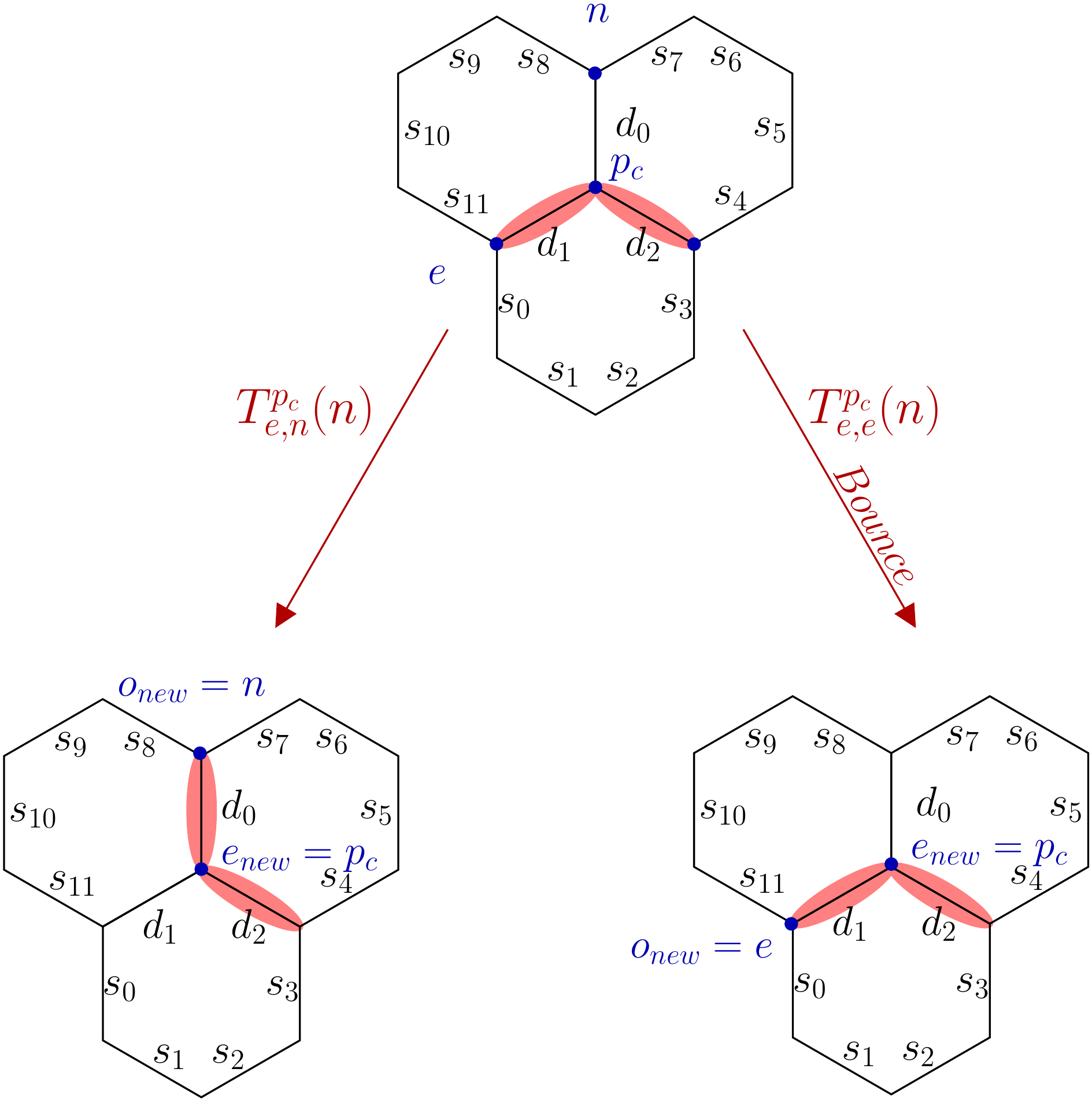}
  \caption{\label{2by2} The pivot step of the DEP worm construction when one arrives at a central pivot site $p_c$ from an entry site $e$, and there are two dimers touching $p_c$: We can exit to the neighbour $n$, which is the neighbour not connected to the central pivot site $p_c$ by a dimer, or bounce back by exiting through the entry site $e$. If we choose to exit through the neighbour $n$, we flip the dimer on the link $\langle p_c e\rangle$ to the link $\langle p_c n \rangle$. We then move along this dimer to reach our new overlap site $o_{new} \equiv n$. If we choose to bounce back, the entry site $e$ becomes our new overlap site $o_{new}$. The central pivot site $p_c$ now becomes our new entry site $e_{new}$ in either case. The probability table $T$ is used to determine which option is chosen. On the dual honeycomb lattice, knowledge of the local dimer configuration consisting of dimer states from $s_0$ to $s_{11}$ and $d_0$, $d_1$ and $d_2$ suffices to calculate $T$ when the interactions extend upto next-next-nearest neighbours on the triangular lattice.
}
\end{figure}

For interactions that extend up to next-next-nearest neighbours on the triangular lattice,
the three weights $W^{v}_i$ that enter these equations differ from each
other only due to factors that depend on the dimer state, $d_0, d_1, d_2$ of the the three links emanating from $v$ and the twelve dual links surrounding $v$, whose dimer
state has been denoted $s_0$, $s_1$ \dots $s_{11}$ in Fig.~\ref{myopic_3by3}. This feature allows us
to tabulate all possible local environments of $v$, and analyze these constraint equations in advance to determine and tabulate the $A^{v}$ (and thence determine $M^{v}$) in advance. In practice, if the weights permit it, we use the ``zero-bounce'' solution given in Ref.~\onlinecite{Syljuasen_Sandvik} and Ref.~\onlinecite{Syljuasen}, else the ``one-bounce'' solution given there.

Having reached the exit $x^{(1)}$ of the vertex $v^{(1)}$ in this manner, we now need to choose the next vertex $v^{(2)}$ which we will enter next from this site $x^{(1)}$. This is the myopic
part of our procedure: {\em This next vertex $v^{(2)}$ is randomly chosen to be one
of the two other neighbours of $x^{(1)}$ (other than the previous vertex $v^{(1)}$) with probability $1/2$ each}(Fig.~\ref{myopic_triangular}). After making this choice,
$x^{(1)}$ becomes the entry site $n^{(2)}$  for this next vertex $v^{(2)}$, and the previous probabilistic procedure is repeated at this next vertex $v^{(2)}$ in order to choose the next exit site $x^{(2)}$ from which we will exit $v^{(2)}$.

In this manner, we go through a sequence of vertices until the exit site  $x^{(k)}$ 
of the $k^{\mathrm{th}}$ vertex equals the start site $o$. When this
happens, one obtains a new dimer configuration which again has either
one dimer touching each honeycomb site, or three dimers touching a honeycomb site.
This new dimer configuration can be accepted with probability one since
our procedure builds in detailed balance with respect to the Boltzmann weight of the generalized dimer model.

It is straightforward to prove this explicitly using the notation we have
developed above. To this end, we first note that the forward
probability for constructing a particular worm to go from an initial configuration
${\mathcal C}_i$ to a final configuration ${\mathcal C}_f$ takes on the product form
\begin{eqnarray}
P({\mathcal C}_i \rightarrow {\mathcal C}_f) &=& \frac{1}{3} \times T^{v^{(1)}}_{n^{(1)}   x^{(1)}} \frac{1}{2} T^{v^{(2)}}_{n^{(2)}   x^{(2)}} \dots \frac{1}{2} T^{v^{(k)}}_{n^{(k)}   x^{(k)}} \nonumber \\
&&
\end{eqnarray}
while the reverse probability takes the form
\begin{eqnarray}
P({\mathcal C}_f\rightarrow {\mathcal C}_i) &=& \frac{1}{3} \times T^{v^{(k)}}_{x^{(k)}   n^{(k)}} \frac{1}{2} T^{v^{(k-1)}}_{x^{(k-1)}   n^{(k-1)}} \dots \frac{1}{2} T^{v^{(1)}}_{x^{(1)}   n^{(1)}} \nonumber \\
&&
\end{eqnarray}
As noted earlier, while the weights $w$ that appear
in the intermediate steps of the construction are computed ignoring the violation of
the generalized dimer constraints at two sites, the initial and final weights
$w^{v^{(1)}}_{n^{(1)}} $ and  $w^{v^{(k)}}_{x^{(k)}} $ have no such caveats associated with them.
Indeed, we have 
\begin{eqnarray}
w^{v^{(1)}}_{n^{(1)}} & \equiv & w({\mathcal C}_i) \;, 
\end{eqnarray}
 the physical Boltzmann weight of the initial configuration, while 
\begin{eqnarray}
w^{v^{(k)}}_{x^{(k)}} & \equiv & w({\mathcal C}_f) \; ,
\end{eqnarray}
 the physical Boltzmann weight of the final configuration.

Now, since our choice of transition probabilities obeys
\begin{eqnarray}
w^{v^{(p)}}_{n^{(p)}}  T^{v^{(p)}}_{n^{(p)}   x^{(p)}} &=& w^{v^{(p)}}_{x^{(p)}}  T^{v^{(p)}}_{x^{(p)}   n^{(p)}} 
\end{eqnarray}
for all $p=1,2\dots k$, and since
\begin{eqnarray}
w^{v^{(p)}}_{x^{(p)}} &\equiv& w^{v^{(p+1)}}_{n^{(p+1)}} 
\end{eqnarray}
for all $p=1,3 \dots k-1$, we may write the following chain of equalities 
\begin{eqnarray}
&&w({\mathcal C}_i)P({\mathcal C}_i \rightarrow {\mathcal C}_f)  =\nonumber \\
&&=w^{v^{(1)}}_{n^{(1)}} \times \frac{1}{3} T^{v^{(1)}}_{n^{(1)}   x^{(1)}} \frac{1}{2} T^{v^{(2)}}_{n^{(2)}   x^{(2)}}  \dots \frac{1}{2}T^{v^{(k)}}_{n^{(k)}   x^{(k)}} \nonumber \\
&&= \frac{1}{3} \times T^{v^{(1)}}_{x^{(1)}   n^{(1)}} w^{v^{(1)}}_{x^{(1)}} \frac{1}{2} T^{v^{(2)}}_{n^{(2)}   x^{(2)}} \dots \frac{1}{2}T^{v^{(k)}}_{n^{(k)}   x^{(k)}} \nonumber \\
&&=  \frac{1}{3} \times T^{v^{(1)}}_{x^{(1)}   n^{(1)}} w^{v^{(2)}}_{n^{(2)}} \frac{1}{2} T^{v^{(2)}}_{n^{(2)}   x^{(2)}} \dots \frac{1}{2}T^{v^{(k)}}_{n^{(k)}   x^{(k)}} \nonumber \\
&&= \frac{1}{3} \times T^{v^{(1)}}_{x^{(1)}   n^{(1)}}  \frac{1}{2} T^{v^{(2)}}_{x^{(2)}   n^{(2)}} w^{v^{(2)}}_{x^{(2)}} \dots \frac{1}{2}T^{v^{(k)}}_{n^{(k)}   x^{(k)}} \nonumber \\
&& \dots \nonumber \\
&&= w^{v^{(k)}}_{x^{(k)}} \times  \frac{1}{3} T^{v^{(k)}}_{x^{(k)}   n^{(k)}} \frac{1}{2} T^{v^{(k-1)}}_{x^{(k-1)}   n^{(k-1)}} \dots \frac{1}{2} T^{v^{(1)}}_{x^{(1)}   n^{(1)}} \nonumber \\
&&=w({\mathcal C}_f)P({\mathcal C}_f\rightarrow {\mathcal C}_i)
\end{eqnarray}
Thus, our procedure explicitly obeys detailed balance, and this myopic worm construction provides a rejection-free update scheme that can effect large changes in the configuration of a generalized honeycomb lattice dimer model with one or three dimers touching each honeycomb site.

\begin{figure}[t]
  \includegraphics[width=8.4cm]{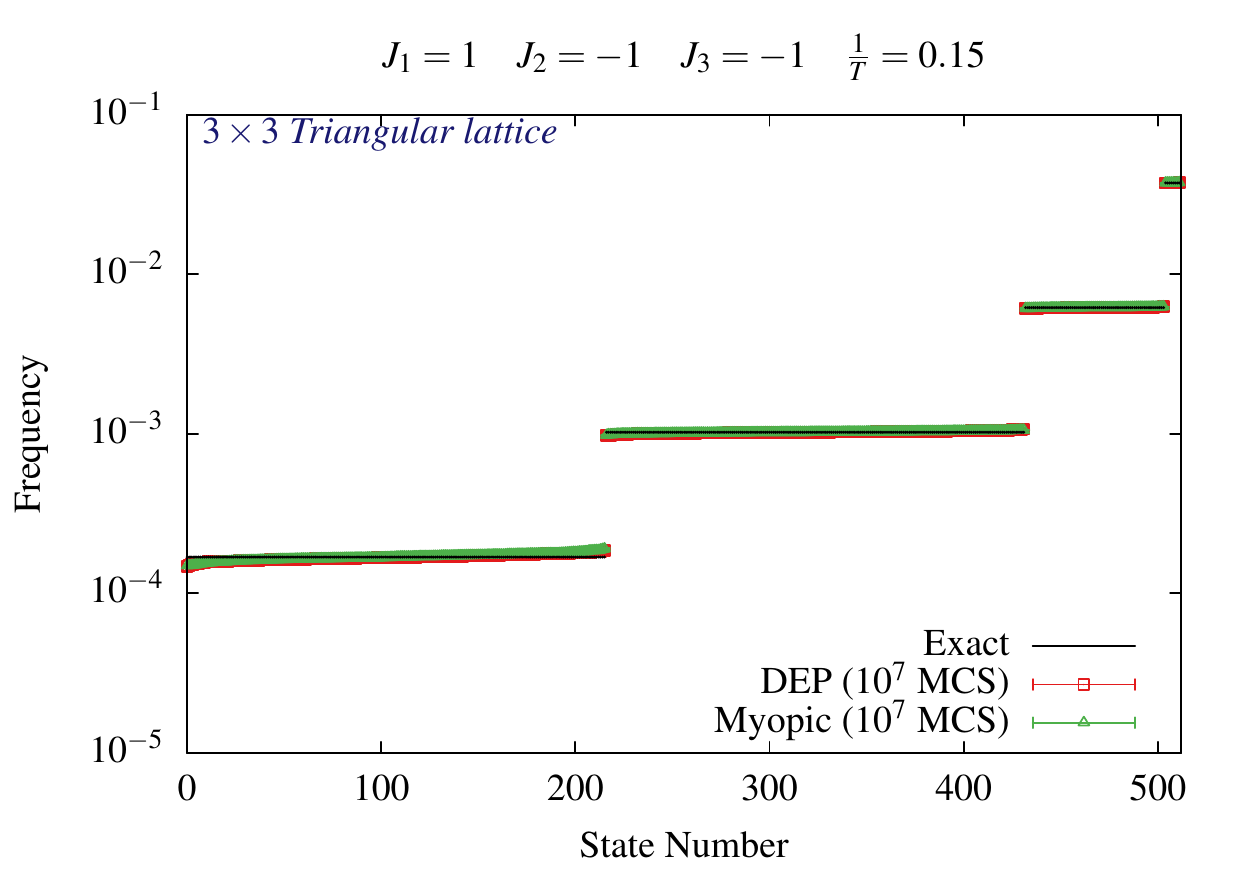}
  \caption{\label{T_freq_allj} The frequency of configurations accessed by a Monte Carlo simulation consisting of $10^7$ Monte Carlo steps (as defined in Sec.~\ref{Performance}) for a $3 \times 3$ triangular lattice Ising model with interactions extending to next-next-nearest neighbours. There are 512 possible unique configurations. The frequencies measured for the DEP worm algorithm and the myopic worm algorithm are seen to agree well with the
predictions of equilibrium Gibbs-Boltzmann statistics, thus establishing the validity of both
algorithms.}
\end{figure}
To translate back into spin language, we need to take care of one additional
subtlety: Although the procedure outlined above gives us a rejection-free nonlocal
update for the generalized dimer model with Boltzmann weight inherited from
the original spin system, we cannot translate this directly into a rejection-free
nonlocal update for the original spin system since we are working on a torus with
periodic boundary conditions for the spin system. The reason has to do with the
fact that the periodic boundary conditions of the spin system translate to
a pair of global constraints: In every valid dimer configuration obtained
from a spin configuration with periodic boundary conditions, the number of
empty links crossed by a path looping around the torus along $\hat{x}$ or $\hat{y}$
must be even, since the absence of a dimer on a dual link perpendicular to
a given bond of the spin system implies that the spins connected by that bond
are antiparallel. This corresponds to constraints on the global winding numbers
of the corresponding dimer model (see Ref.~\onlinecite{Patil_Dasgupta_Damle} for
a definition specific to the honeycomb lattice dimer model), which must
be enforced by any Monte Carlo procedure. Note that these constraints are
on the parity of these winding numbers, which are in any case only defined modulo $2$
unless one is at $T=0$.

Therefore, to convert this rejection-free myopic worm update procedure for dimers
into a valid update scheme for the original spin system, we test the winding numbers (modulo $2$) of the new dimer configuration to see if it satisfies these two global constraints. If the answer
is yes, we translate the new dimer configuration back into spin language by
choosing the spin at the origin to be up or down with probability $1/2$ and
reconstructing the remainder of the spin configuration from the positions of the dimers.
If, however, the new dimer configuration is in an illegal winding sector, we repeat
the previous spin configuration in our Monte Carlo chain.

This procedure generalizes readily to the Kagome lattice Ising antiferromagnet
with interactions extending up to next-next-nearest neighbour spins. Since most
of the required generalizations are self-evident, we merely point out some
of the key differences here. Our myopic worm update procedure
now begins with a randomly chosen six-coordinated site as the start site $o$.
With probabilities $1/6$ each, we choose one of its neighbours as the first
vertex site $v^{(1)}$ (Fig.~\ref{start_myopic_kagome}). The start site $o$ thus becomes the entry site $n^{(1)}$ from
which we enter the first vertex $v^{(1)}$. The choice of the first exit
site $x^{(1)}$ via which we exit the first vertex is again dictated by a three-by-three
probability table(Fig.~\ref{myopic_3by3}). 

For any vertex $v$, this probability table is determined by solving detailed balance equations completely analogous to the ones displayed earlier for the honeycomb lattice case. In the dice lattice case, the three weights $W^{v}_{i}$ depend on the dimer states
$s_0$, $s_1$ \dots $s_5$ of the six dual links shown in Fig.~\ref{myopic_3by3}, and on the dimer states $d_0, d_1, d_2$ of the three links emanating from $v$. Therefore, we are again in a position to solve these equations for all possible local environments of $v$ and tabulate these solutions for repeated use during the worm construction.

As in the honeycomb case, having arrived at $x^{(1)}$, we choose the next vertex $v^{(2)}$ 
in a myopic manner: Without regard to the local dimer configuration, we randomly
pick, with probability $1/5$ each, one of the other neighbours (other than $v^{(1)}$)
of $x^{(1)}$ as the next vertex $v^{(2)}$ (Fig.~\ref{myopic_kagome}). $x^{(1)}$ now becomes the entry site $n^{(2)}$
from which we enter this second vertex $v^{(2)}$. The exit $x^{(2)}$ is again
chosen from the pre-tabulated probability table, and the process continues until
the $k^{\mathrm{th}}$ exit $x^{(k)}$ equals then start site $o$.

Clearly, our earlier proof of detailed balance goes through unchanged, and this myopic worm construction again gives a rejection-free way of
updating the dual dimer model in accordance with detailed balance. To translate this into an update
scheme for the original spin model, we must again check that the new dimer
configuration is in a legal winding sector, and if the new configuration is in
an illegal winding sector, we must repeat the original spin configuration in our
Monte Carlo chain.

\subsection{DEP worm algorithm}
\begin{figure}[t]
  \includegraphics[width=8.4cm]{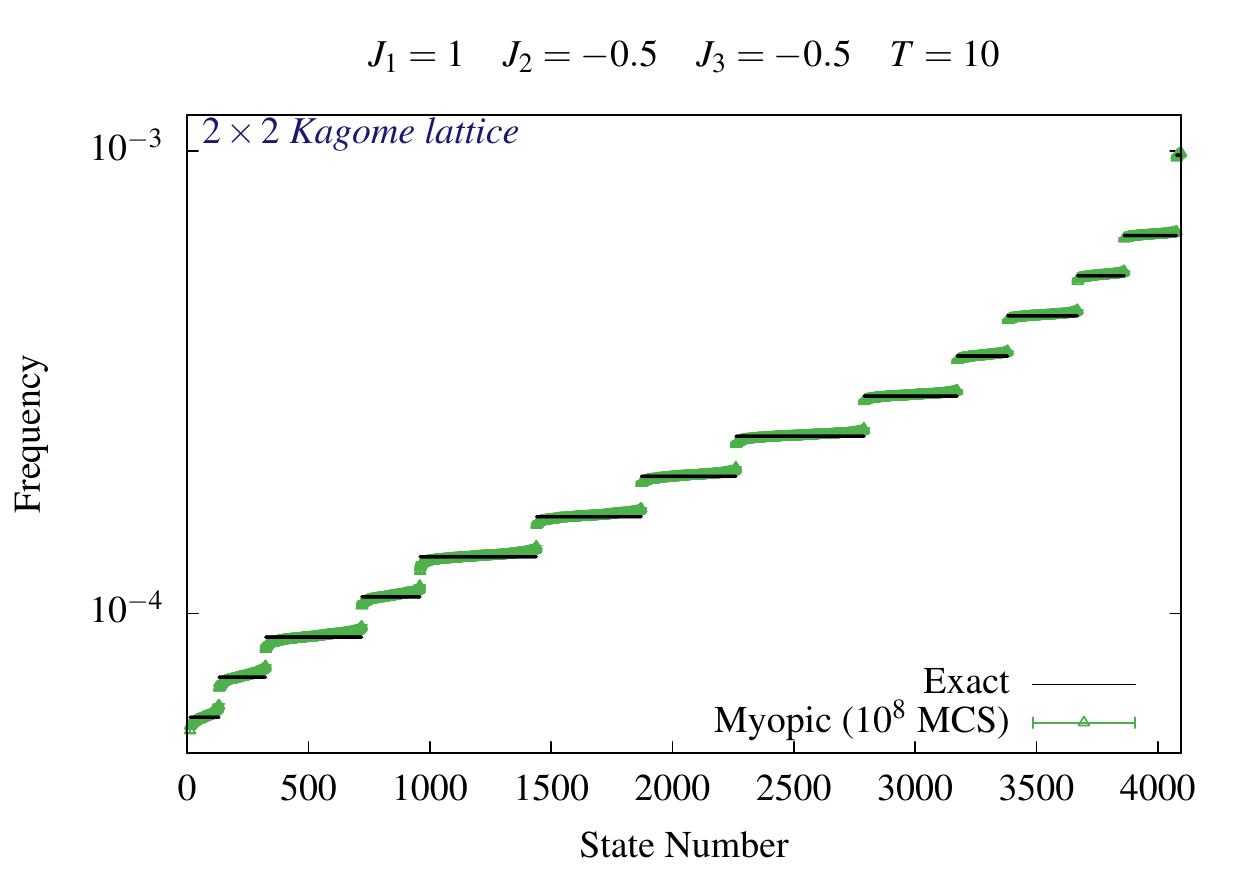}
  \caption{\label{K_freq_allj} The frequency of configurations accessed by a Monte Carlo simulation consisting of $10^8$ Monte Carlo steps (as defined in Sec.~\ref{Performance}) for a $2 \times 2$ Kagome lattice Ising model with interactions extending to next-next-nearest neighbours. There are 4096 possible unique configurations. The frequencies measured for the myopic worm algorithm are seen to agree well with the
predictions of equilibrium Gibbs-Boltzmann statistics, thus establishing the validity of the
algorithm.
}
\end{figure}
The other strategy we have developed is specific to the honeycomb lattice
dimer model that is dual to the triangular lattice Ising antiferromagnet. Since it
involves deposition, evaporation, and pivoting of dimers, we dub this
the DEP worm algorithm. The DEP worm construction begins by choosing
a random start site $s$. The subsequent worm construction consists of so-called
``overlap steps'' and ``pivot steps''. Pivot steps are carried out when
one reaches a ``central'' ``pivot site'' from a neighbouring ``entry site'', while overlap steps are carried out when one reaches a central ``overlap site'' from a neighbouring
entry site. Details of some of the subsequent pivot steps
in this construction depend on  whether the randomly chosen start site $s$ is touched by three dimers or by one dimer, {\em i.e.} if the corresponding
triangle is defective or minimally frustrated. Therefore we describe these two
branches of the procedure separately, but use a unified notation
so as to avoid repetition of the aspects that do not depend on the branch chosen.

\subsubsection{Branch I}

Let us first consider the case when the randomly chosen start site $s$ is touched by exactly one dimer(Fig.~\ref{start_branch1}). In this case,
we move along the dimer touching $s$ to its other end. The site
at the other end of this dimer becomes our first central site $c^{(1)}$, at which
we must now employ a pivot step with $c^{(1)}$ as the first pivot site. For
the purposes of this first pivot step, the start site $s$ becomes
the entry site $e^{(1)}$ from which we have arrived at this pivot site $c^{(1)}$ by walking along this dimer.

Before proceeding further, it is useful to elucidate the nature of a general pivot
move encountered in our algorithm: In a pivot step, after one
arrives at the central pivot site $p_{\mathrm{c}}$ from an entry site $e$ (as we will
see below, $e$ could
be the previous overlap site $o_{\mathrm{old}}$, or a previous pivot site $p_{\mathrm{old}}$) by moving along a dimer connecting $e$ to $p_{\mathrm{c}}$, the subsequent protocol depends on whether there is exactly one dimer (Fig.~\ref{loop}) touching $p_{\mathrm{c}}$ or three (Fig.~\ref{kill}).
In the first case, one pivots the dimer touching $p_{\mathrm{c}}$, so that it now
lies on link $\langle p_{\mathrm{c}} o_{\mathrm{new}} \rangle$ instead of link  $\langle p_{\mathrm{c}} e \rangle$(Fig.~\ref{loop}). Here, $o_{\mathrm{new}}$ is one of the neighbours of $p_{\mathrm{c}}$, chosen
using the element $R^{p_{\mathrm{c}}}_{e,o_{\mathrm{new}}}$  of a three-by-three probability table $R^{p_{\mathrm{c}}}_{\alpha, \beta}$ (where $\alpha$ and $\beta $ range over the three neighbours
of the central site $p_{\mathrm{c}}$, and the full structure of this table is specified at the end of this
discussion). Note that in some cases, it is possible for  $o_{{\mathrm{new}}} = e$ with nonzero probability, if the corresponding diagonal entry of the table is nonzero. After this is done, 
the next step in the construction will be an overlap step, with $o_{\mathrm{new}}$ being the central overlap site and $p_{{\mathrm{c}}}$ playing the role of the new entry
site $e_{\mathrm{new}}$ from which we have arrived at this central overlap site. The structure
of a general overlap step is specified below, after describing the pivot move in
the second case, {\em i.e.} with three dimers touching the central pivot site.

If the central pivot site $p_{\mathrm{c}}$ in a pivot step has three dimers connecting
it  (Fig.~\ref{kill}) to its three neighbours  $n_1$, $n_2$, and $e$ (where $e$ is the entry
site from which we arrived at the central pivot site $p_{\mathrm{c}}$),  we choose one
out of three alternatives using a different probability table $K^{c_s}_{\alpha, \beta}(n_p)$, where $\alpha$ and $\beta$
range over all neighbours of a central site $c_s$ and $n_p$ is a particular
privileged neighbour of $c_s$ (in the case being described here, $c_s= p_{\mathrm{c}}$
and $n_p=e$): With probabilities $K^{p_c}_{e, n_1}(e)$ 
and  $K^{p_c}_{e, n_2}(e)$ drawn respectively from this table, we may delete the dimer on link $\langle p_{\mathrm{c}} e \rangle$ and reach either $n_1$ or $n_2$, and the next step would
then be a pivot step, with the neighbour thus
reached now playing the role of the new central pivot site $p_{\mathrm{new}}$ and $p_{\mathrm{c}}$ playing the role of the new entry site $e_{\mathrm{new}}$ from which we have reached this new central pivot site. On the other hand, we may 
``bounce'' with probability $K^{p_c}_{e, e}(e)$ , {\em i.e.} we simply return from $p_{\mathrm{c}}$ to $e$ without deleting any of the three dimers touching $p_{\mathrm{c}}$; in this case, the next step will be an overlap step, with $e$ as the new central overlap site $o_{\mathrm{new}}$, and $p_{\mathrm{c}}$ will play the role of the new entry site $e_{\mathrm{new}}$ from which we have reached this new overlap site (Fig.~\ref{kill}). Note that elements of this
table $K^{c_s}_{\alpha, \beta}(n_p)$ with $\alpha \neq n_p$ never play any role in the
choices made at this kind of pivot step. As we will see below, these elements
of the table in fact determine the
choices made at a general overlap step in a way that preserves local detailed balance.

Returning to our construction, if the central pivot site $c^{(1)}$ was of the second type
and we did not bounce, we would reach a new central pivot site $c^{(2)}$ (with $c^{(1)}$ now
becoming the entry site $e^{(2)}$ from which we reach this new pivot site), and
we would perform another pivot step as described above. If on the other
hand, the central pivot site $c^{(1)}$ was of this first type or if it was of
the second type and we bounced, the next step will be an overlap step
with a new central overlap site $c^{(2)}$. Since we would have reached
 $c^{(2)}$ by moving along a dimer connecting it to $c^{(1)}$, $c^{(1)}$ will  play the role of the new entry site $e^{(2)}$ for this overlap step (in the bounce case,  $c^{(2)} = e^{(1)}$). Having reached the central overlap site $c^{(2)}$ from entry site $e^{(2)}$ in this way, we must
employ an overlap step. 

Before proceeding with our construction, let us first elucidate the structure of choices
at an overlap step after we have reached a central overlap site $o_c$ from an entry site $e$. $e$ could be the previous 
pivot site $e=p_{\mathrm{old}}$ if the previous step had been a pivot step (as in the example above) or it could be a previous overlap
site $e=o_{\mathrm{old}}$ if the previous step had also been
an overlap step (we will see below that this is also possible).
In either case, at a general overlap step, one arrives at the overlap site $o_c$
from entry site $e$ along one of the two dimers touching $o_c$. Thus, one neighbour
of $o_c$, suggestively labeled $o_{\mathrm{new}}$, is {\em not} connected to
the central overlap site $o_c$ by a dimer, while the other two neighbours are connected to
$o_c$ by dimers. One of the latter pair of neighbours is of course the entry site $e$ from which we arrived at $o_c$, while we suggestively label the other as $p_{\mathrm{new}}$.

At such an overlap step, one always has two options to choose from, whose
probabilities are given as follows by entries of the probability table $K$ introduced earlier(Fig~\ref{make}): One option is to deposit, with
probability $K^{o_c}_{e, o_{\mathrm{new}}}(o_{\mathrm{new}})  $, an additional dimer on the originally empty link $\langle o_c o_{\mathrm{new}} \rangle$ emanating from $o_{c}$. If we do this,
$o_{\mathrm{new}}$ becomes the new overlap site, which we have entered from $o_c$,
which becomes the new entry site $e_{\mathrm{new}}$, and the next step 
will again be an overlap step. The second option, chosen with
probability  $K^{o_c}_{e, p_{\mathrm{new}}} (o_{\mathrm{new}})$, is that we move along the second dimer touching $o_{c}$ to the other neighbour $p_{\mathrm{new}}$, which is connected to $o_{c}$ by this second dimer. If we do this,
$p_{\mathrm{new}}$ becomes the new pivot site, which we enter from site $o_c$,
which becomes the new entry site $e_{\mathrm{new}}$, and the next step will
be a pivot step. As we will see below, the fact that the table $K$ that fixes the probabilities for choosing between
these two options is the same as the one
used in a pivot step (when the pivot site has three dimers touching it) is crucial in
formulating and satisfying local detailed balance conditions that guarantees a
rejection-free worm update.

Returning again to our construction, we employ this procedure to
carry out an overlap step when we reach the overlap site $o^{(2)}$ from
entry site $e^{(2)}$. Clearly this process continues until we encounter
the start site $s$ as the new overlap site in the course of our worm construction.
When this happens, we obtain a new dimer configuration that satisfies
the generalized dimer constraint that each site be touched by one or three dimers.

\subsubsection{Branch II}

\begin{figure}[t]
  \includegraphics[width=8.4cm]{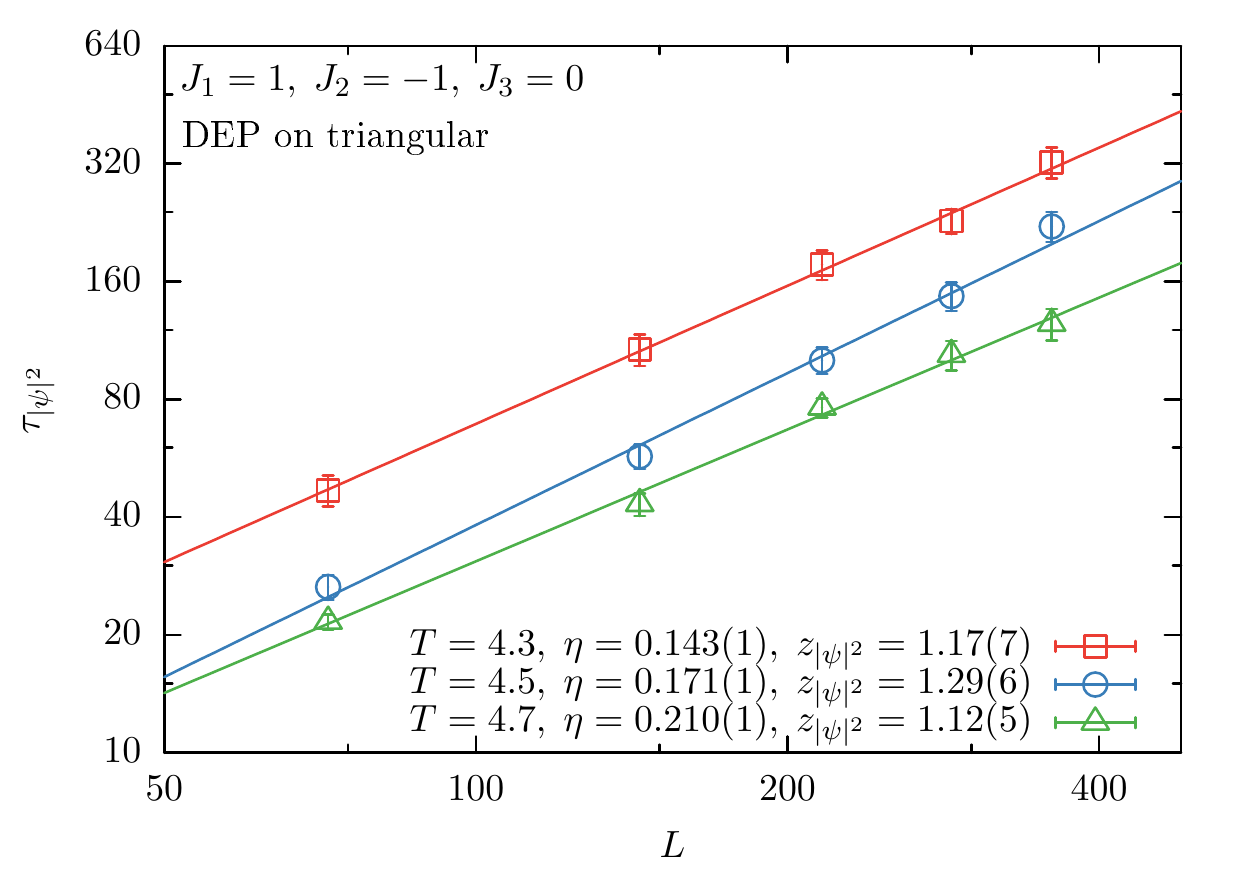}
  \caption{\label{T_gen_ac_psi_j1_j2} The lattice size $L$ dependence of autocorrelation time of the three sublattice order parameter $|\psi|^2$ in Monte Carlo simulations using the DEP worm algorithm on the triangular lattice with nearest neighbour antiferromagnetic and next-nearest neighbour ferromagnetic interactions. The dynamical exponent $z$ is extracted by fitting to the functional form $cL^z$ at three temperatures at which the system is in the power-law ordered critical phase. Also shown is the anomalous exponent $\eta$ of the power law three-sublattice order at the corresponding temperatures.
}
\end{figure}

\begin{figure}[t]
  \includegraphics[width=8.4cm]{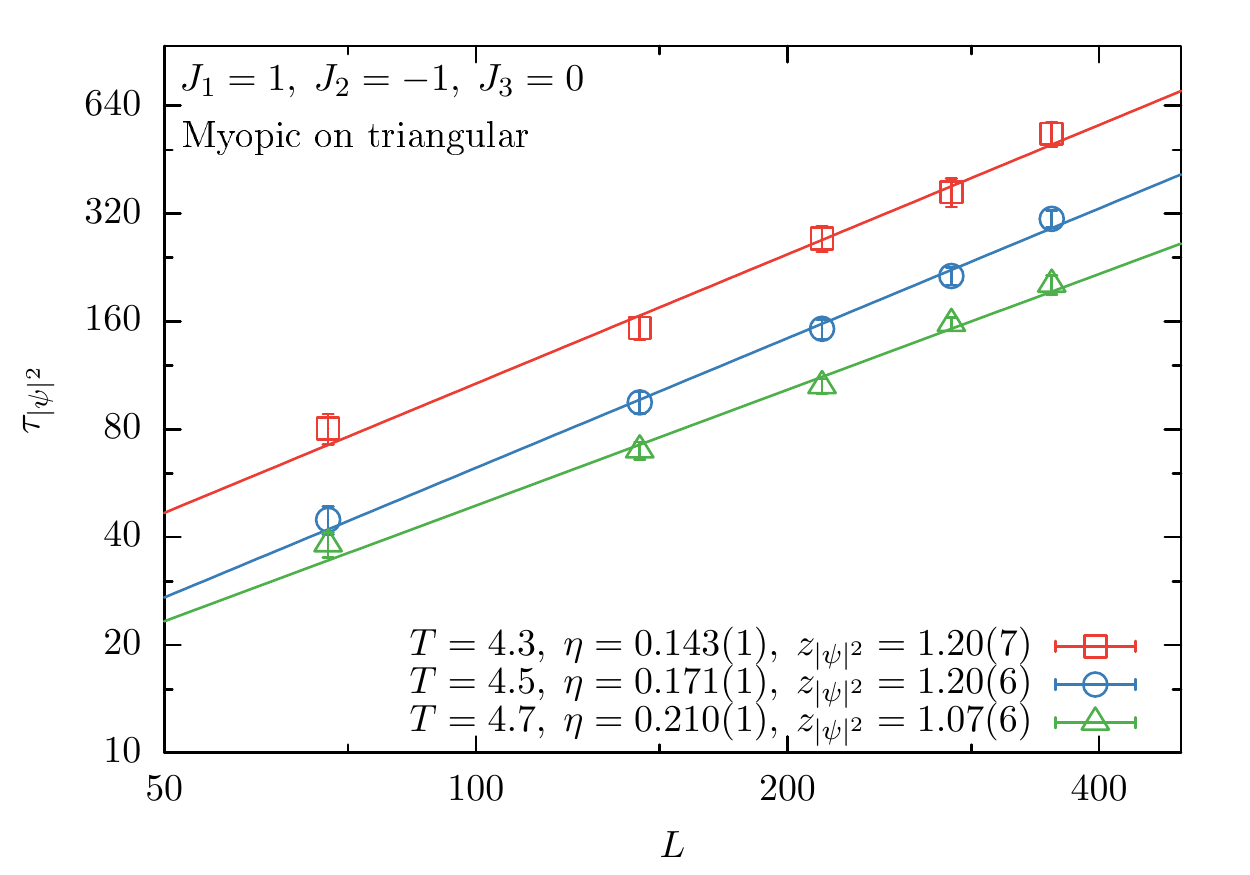}
  \caption{\label{T_myo_ac_psi_j1_j2} The lattice size $L$ dependence of autocorrelation time of the three sublattice order parameter $|\psi|^2$ in Monte Carlo simulations using the myopic worm algorithm on the triangular lattice with nearest neighbour antiferromagnetic and next-nearest neighbour ferromagnetic interactions. The dynamical exponent $z$ is extracted by fitting to the functional form $cL^z$ at three temperatures at which the system is in the power-law ordered critical phase. Also shown is the anomalous exponent $\eta$ of the power law three-sublattice order at the corresponding temperatures.
}
\end{figure}

\begin{figure}[t]
  \includegraphics[width=8.4cm]{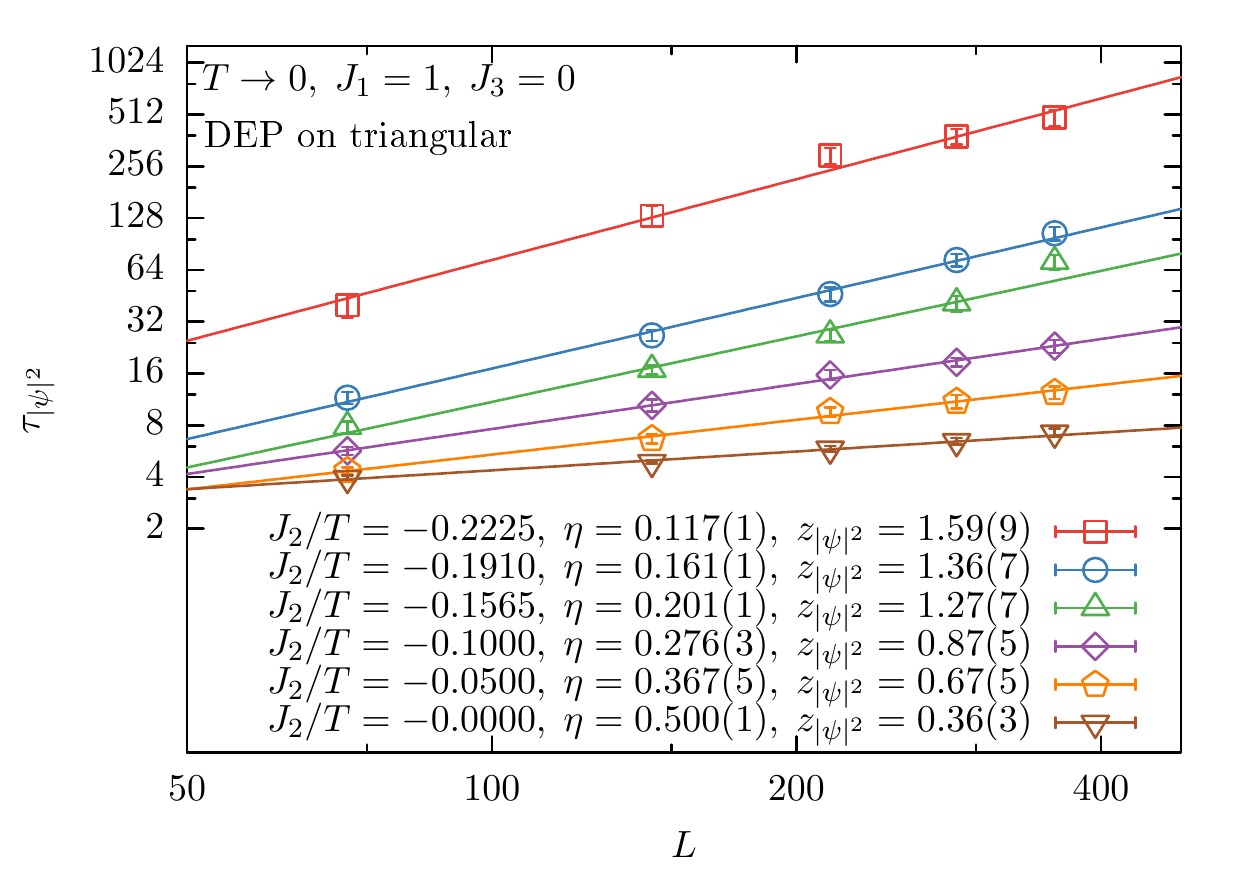}
  \caption{\label{T_gen_ac_psi_onlyj2} The lattice size $L$ dependence of autocorrelation time of the three sublattice order parameter $|\psi|^2$ in Monte Carlo simulations using the DEP worm algorithm on the triangular lattice with nearest neighbour antiferromagnetic and next-nearest neighbour ferromagnetic interactions. The dynamical exponent $z$ is extracted by fitting to the functional form $cL^z$ at six values of $J_2/T$ at which the system is in the power-law ordered critical phase in the zero temperature limit $T \rightarrow 0$. Also shown is the anomalous exponent $\eta$ of the power law three-sublattice order at the corresponding points in the zero temperature phase diagram.
}
\end{figure}

Let us now consider the case when the randomly chosen start site $s$ is touched by three dimers. In this case, we move along one of the three dimers touching $s$ to its other end (with probability $1/3$ each), so that the site at the other end becomes the central pivot site $c^{(1)}$ for a pivot step, and the start site $s$ becomes the entry site $e^{(1)}$ from which
we have entered this central pivot site (Fig~\ref{start_branch2}). We now implement the protocol
for a pivot step (as described in {\em Branch I}) to reach a new central site $c^{(2)}$. If the next step turns out to be a pivot step, $c^{(2)}$ plays the role
of a central pivot site, whereas it becomes the central overlap site if the next step
is an overlap step. In either case, $c^{(1)}$ becomes the entry site $e^{(2)}$ for this next step. In this manner, we continue until we reach the start site $s$ as the new central overlap site
$c^{(k)}$ for an overlap step. When this happens, the worm construction
ends after these $k$ steps, since the start site again has three dimers touching it, and we thus obtain a new dimer configuration that satisfies the generalized dimer constraint that each site be touched by one or three dimers. 

The only additional feature introduced in {\em Branch II} is that one could in principle
reach the start site $s$ as the central pivot site of some intermediate pivot step $l$(Fig.~\ref{2by2}). In this
case, the intermediate configuration reached at this $l^{\mathrm{th}}$ step is
not a legal one (since it still has two dimers touching  $s$), and we need to continue
with the worm construction. This is done using a special ``two-by-two'' pivot step(Fig.~\ref{2by2}).
In this two-by-two pivot step, one arrives at the two-by-two pivot site (which will always be the start site in our construction) $p_c$ from an entry site $e$ (as in all other steps, $e$ could be a previous central overlap site $o_{\mathrm{old}}$ or the previous central pivot site $p_{\mathrm{old}}$) by moving along a dimer connecting $e$ to $p_c$. Unlike the usual pivot step, at which there is only one dimer touching
the central pivot site, $p_c$ has a second dimer touching it, which connects $p_c$ to
another neighbour $n_f$. Thus, unlike the usual pivot step, there is just one neighbour of $p_c$, suggestively labeled $o_{\mathrm{new}}$, which is not connected to $p_c$ by a dimer when one arrives at $p_c$ to implement this step. Therefore, our only options are to rotate the dimer which was on link $\langle e p_c \rangle$, to now lie on link $\langle p_c o_{\mathrm{new}} \rangle$, or to bounce.  The probabilities for these
two choices are determined by a probability table $T^{p_c}_{\alpha,\beta}(n_f)$. Here,
 $\alpha$ and $\beta$ are both constrained to not equal $n_f$, making $T^{p_c}_{\alpha,\beta}(n_f)$ a two-by-two matrix.  In either case, $o_{\mathrm{new}}$ chosen in one of these two ways becomes the new central overlap site of the next step, which must be an overlap step, and the process continues.

This new configuration thus obtained upon completing the worm construction initiated
either using {\em Branch I} or {\em Branch II} can now be accepted with probability one if the probabilities with which we carried out each of the intermediate pivot steps and overlap
steps obeyed local detailed balance. Local detailed balance at
a pivot step in which the pivot site is touched by one dimer requires that the probability table $R^{p_{\mathrm{c}}}_{e,o_f}$
obeys the conditions
\begin{eqnarray}
w^{p_c}_e R^{p_c}_{e,o_f} &=& w^{p_c}_{o_f} R^{p_c}_{o_f,e} \; ,
\end{eqnarray}
where the $w^{p_c}_n$ is the Boltzmann weight of the dimer configuration
in which the link $\langle p_c n\rangle$ connecting $p_c$ to one of
its neighbours $n$ is occupied by a dimer and
the other two links emanating from $p_c$ are empty. As in all worm
constructions, these weights are computed ignoring the fact that the generalized
dimer constraint (that each site be touched by exactly one or three dimers) is
violated at two sites on the lattice. These conditions again form a three-by-three set
of constraint equations of the type discussed in Ref.~\onlinecite{Syljuasen_Sandvik}
and Ref.~\onlinecite{Syljuasen}, allowing us to analyze these constraints  and tabulate solutions in advance
for all cases that can
be encountered. If the
weights permit it, we use the ``zero-bounce'' solution given in Ref.~\onlinecite{Syljuasen_Sandvik} and Ref.~\onlinecite{Syljuasen},
else the ``one-bounce'' solution given there.

Local detailed balance at a two-by-two pivot step in which the pivot site is touched by two dimers requires that the probability table $T^{p_c}_{\alpha,\beta}(n_f)$
obeys the conditions
\begin{eqnarray}
w^{p_c}_\alpha(n_f) T^{p_c}_{\alpha,\beta}(n_f) &=& w^{p_c}_{\beta}(n_f) T^{p_c}_{\beta, \alpha}(n_f) \; ,
\end{eqnarray}
where the $w^{p_c}_\alpha(n_f)$ is the Boltzmann weights of the dimer configurations
in which the links $\langle p_c n_f\rangle$ and $\langle p_c \alpha \rangle$ are covered
by dimers and the third link is empty. As always, these weights are computed
ignoring the fact that the dimer constraints are violated at two sites on the dual lattice.
In practice, we tabulate all possible local environments that can arise in
such an update step, and use Metropolis probabilities to tabulate in advance the corresponding entries
of $T^{p_c}(n_f)$.

\begin{figure}[t]
  \includegraphics[width=8.4cm]{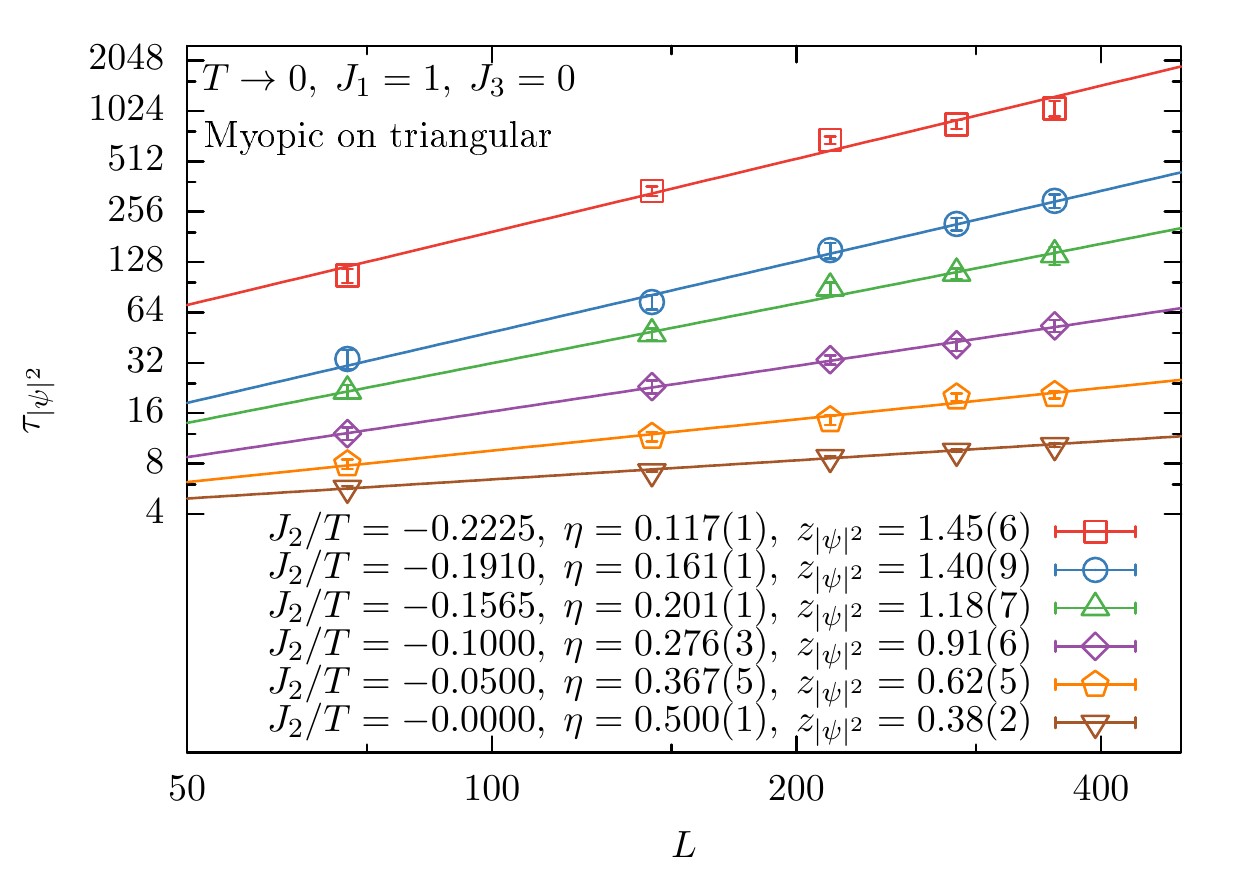}
  \caption{\label{T_myo_ac_psi_onlyj2} The lattice size $L$ dependence of autocorrelation time of the three sublattice order parameter $|\psi|^2$ in Monte Carlo simulations using the myopic worm algorithm on the triangular lattice with nearest neighbour antiferromagnetic and next-nearest neighbour ferromagnetic interactions. The dynamical exponent $z$ is extracted by fitting to the functional form $cL^z$ at six values of $J_2/T$ at which the system is in the power-law ordered critical phase in the zero temperature limit $T \rightarrow 0$. Also shown is the anomalous exponent $\eta$ of the power law three-sublattice order at the corresponding points in the zero temperature phase diagram.
}
\end{figure}

Finally, the constraints imposed by local detailed balance at an overlap step
are essentially intertwined with the local detailed balance constraints that must
be enforced at a pivot step when the pivot site has three dimers touching it. This is
because the deletion of a dimer at such a pivot step is the ``time-reversed'' counterpart
of the process by which an additional dimer is deposited at an overlap step. Indeed,
this is why we have been careful in our discussion above to draw the probabilities
at the pivot step from the same table $K$ as the probabilities that govern the choices
to be made at an overlap step. 

\begin{figure}[t]
  \includegraphics[width=8.4cm]{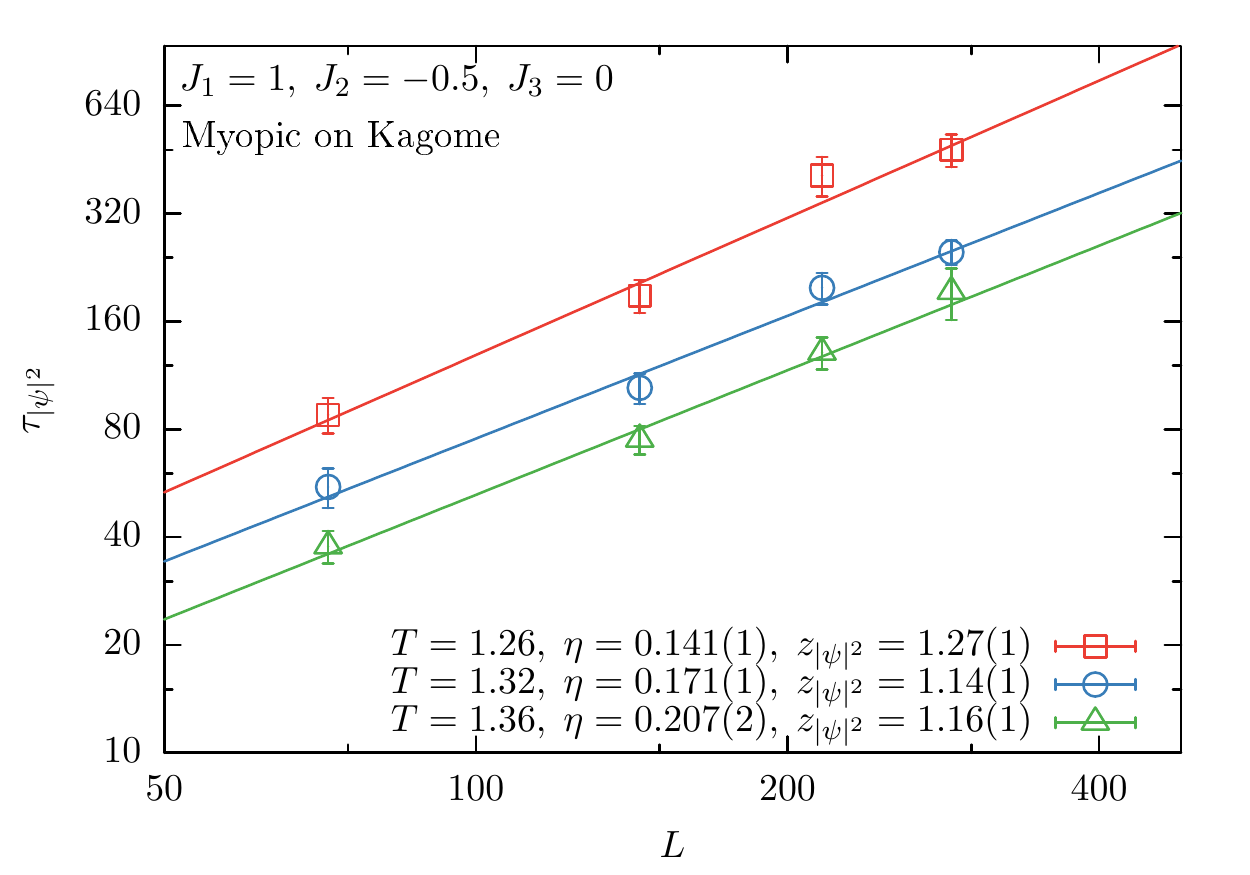}
  \caption{\label{K_myo_ac_psi_j1_j2} The lattice size $L$ dependence of autocorrelation time of the three sublattice order parameter $|\psi|^2$ in Monte Carlo simulations using the myopic worm algorithm on the Kagome  lattice with nearest neighbour antiferromagnetic and next-nearest neighbour ferromagnetic interactions. The dynamical exponent $z$ is extracted by fitting to the functional form $cL^z$ at three temperatures at which the system is in the power-law ordered critical phase. Also shown is the anomalous exponent $\eta$ of the power law three-sublattice order at the corresponding temperatures.
}
\end{figure}

We use a different probability table $K^{c_s}_{\alpha, \beta}(n_p)$ (where $\alpha$ and $\beta$
range over all neighbours of a central site $c_s$ and $n_p$ is a particular
privileged neighbour of $c_s$) to decide on the next course
of action:

We now describe the structure of the table $K^{c_s}_{\alpha, \beta}(n_p)$. Here, $c_s$ is the 
central site, which would be the current pivot site in a pivot step with three dimers
touching the pivot site, or the current overlap site in an overlap step.  $n_p$ is
a ``privileged neighbour'' of $c_s$; in a pivot step, $n_p$ is the entry site from which
we enter the pivot site $c_s$, while in an overlap step, it is the unique neighbour of $c_s$ that is not connected to $c_s$ by a dimer. Clearly, local
detailed balance imposes the following constraints on this probability table $K$:
\begin{eqnarray}
w^{c_s}_\alpha(n_p) K^{c_s}_{\alpha, \beta}(n_p) &= & w^{c_s}_\beta(n_p) K^{c_s}_{\beta, \alpha}(n_p) \; . 
\end{eqnarray}
Here both $\alpha$ and $\beta$ can be either the site $n_p$ or the two
other neighbours $n_1$ and $n_2$ of the central site $c_s$. $w^{c_s}_{n_1}(n_p) = w^{c_s}_{n_2}(n_p) $ denotes the weight of the
configuration with both links $\langle c_s n_1\rangle$ and $\langle c_s n_2\rangle$ covered
by a dimer and the link $\langle c_s n_p \rangle $ unoccupied by a dimer. On the
other hand, $w^{c_s}_{n_p}(n_p)$ denotes the weight of the configuration
in which all three links $\langle c_s n_1\rangle$, $\langle c_s n_2\rangle$ 
and  $\langle c_s n_{p}\rangle$ are covered by dimers. As before, these weights are computed ignoring the fact that the generalized
dimer constraint (that each site be touched by exactly one or three dimers) is
violated at two sites on the lattice.

Choices for the tables $R$ and $K$ consistent with these local detailed
balance constraints, can be computed using the same strategy described
in our construction of the myopic worm update. Again, the weights that
enter these constraints on $K$ ($R$) depend only on the dimer states $d_0$, $d_1$, and $d_2$ of the three links emanating from the central site $c_s$ (pivot site $p_c$),
and the dimer states $s_0$, $s_1$ \dots $s_{11}$ of the twelve links surrounding
this site, allowing us to tabulate in advance all possible local environments and the corresponding solutions for $K$ and $R$. The formal proof of detailed balance uses these
local detailed balance constraints to construct a chain of equalities relating
the probabilities for an update step and its time-reversed counterpart in
exactly the same way as the proof given in the previous discussion of
the myopic worm update. Therefore, we do not repeat it here for the present case.

\begin{figure}[t]
  \includegraphics[width=8.4cm]{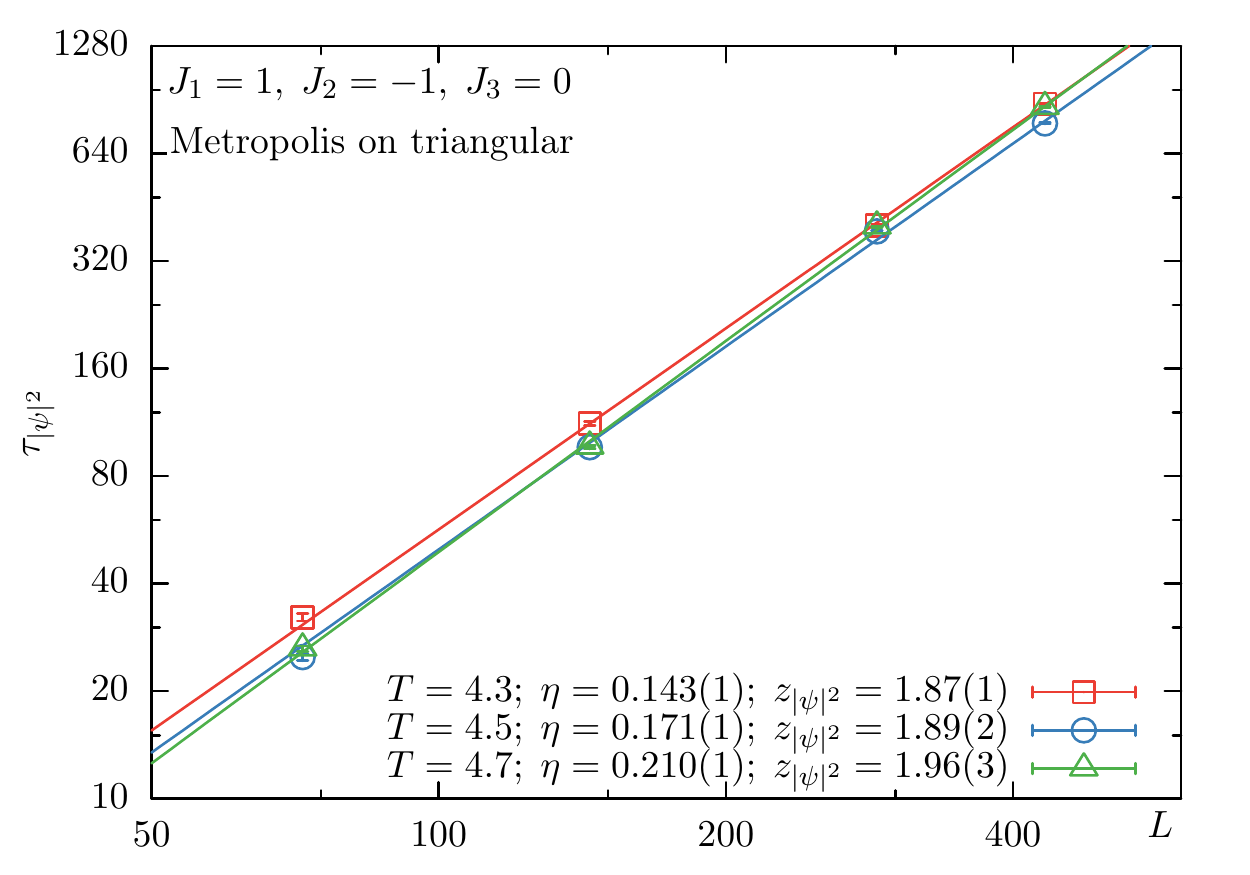}
  \caption{\label{T_met_ac_psi_j1_j2} The lattice size $L$ dependence of autocorrelation time of the three sublattice order parameter $|\psi|^2$ in Monte Carlo simulations using the
Metropolis algorithm on the triangular lattice with nearest neighbour antiferromagnetic and next-nearest neighbour ferromagnetic interactions. The dynamical exponent $z$ is extracted by fitting to the functional form $cL^z$ at three temperatures at which the system is in the power-law ordered critical phase. Also shown is the anomalous exponent $\eta$ of the power law three-sublattice order at the corresponding temperatures.
}
\end{figure}

\begin{figure}[t]
  \includegraphics[width=8.4cm]{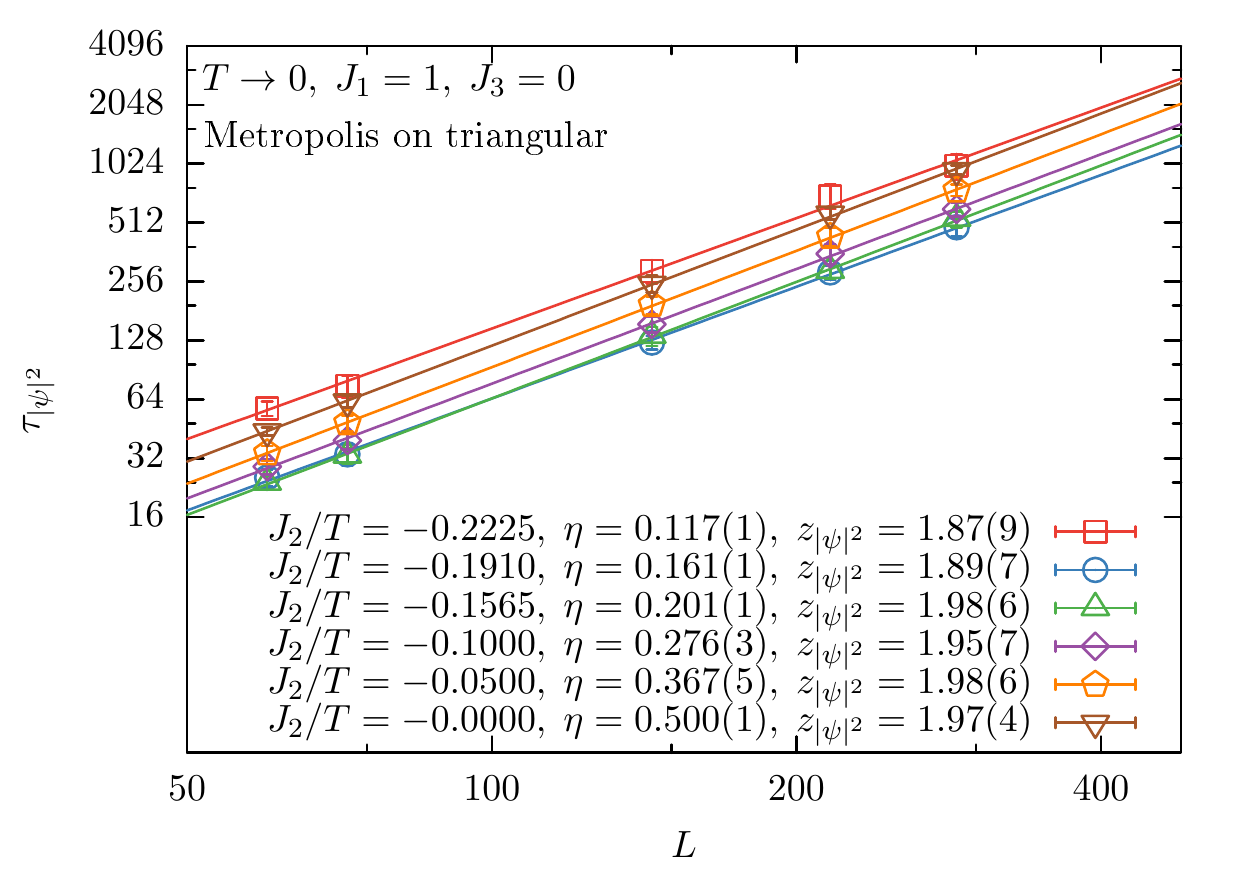}
  \caption{\label{T_met_ac_psi_onlyj2}  The lattice size $L$ dependence of autocorrelation time of the three sublattice order parameter $|\psi|^2$ in Monte Carlo simulations using the Metropolis algorithm on the triangular lattice with nearest neighbour antiferromagnetic and next-nearest neighbour ferromagnetic interactions. The dynamical exponent $z$ is extracted by fitting to the functional form $cL^z$ at six values of $J_2/T$ at which the system is in the power-law ordered critical phase in the zero temperature limit $T \rightarrow 0$. Also shown is the anomalous exponent $\eta$ of the power law three-sublattice order at the corresponding points in the zero temperature phase diagram.
}
\end{figure}

\begin{figure}[t]
  \includegraphics[width=8.4cm]{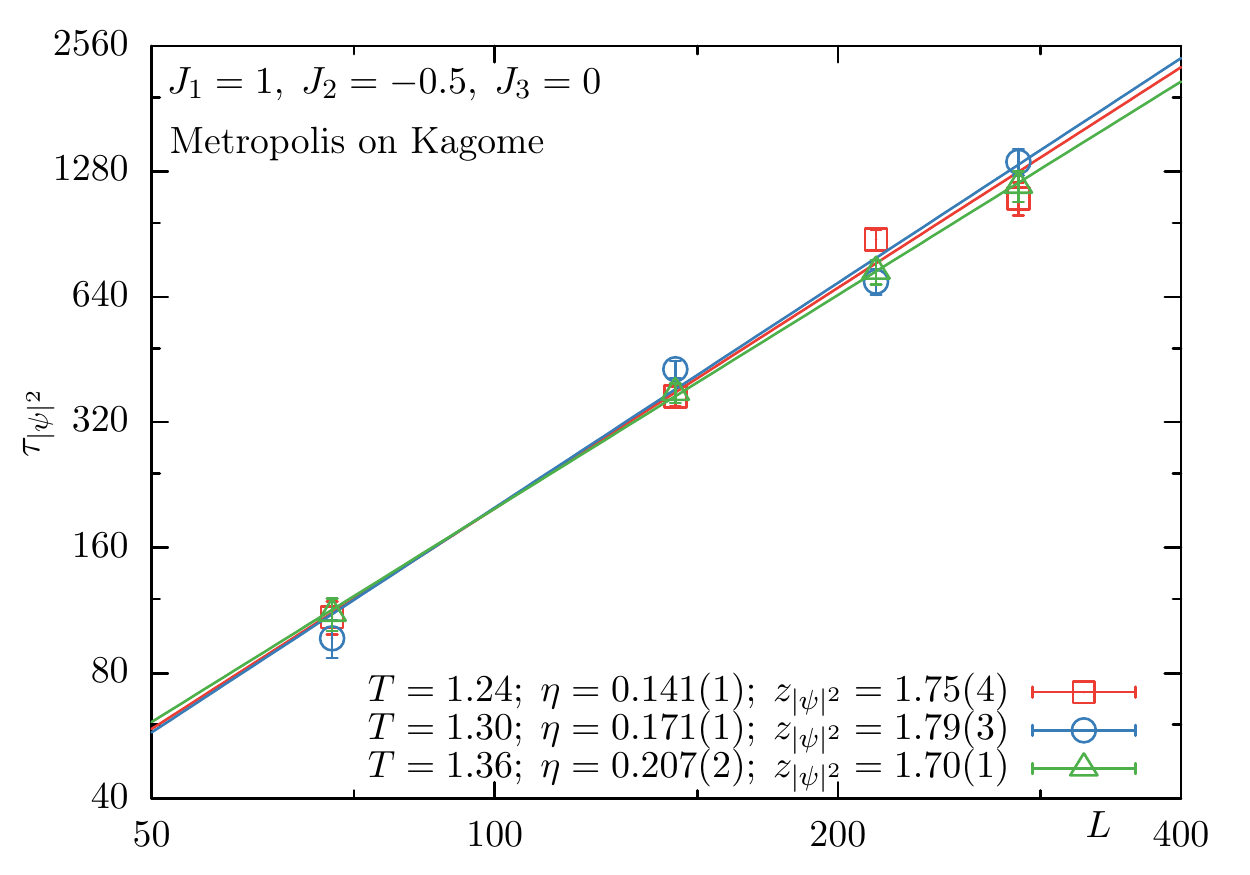}
  \caption{\label{K_met_ac_psi_j1_j2} The lattice size $L$ dependence of autocorrelation time of the three sublattice order parameter $|\psi|^2$ in Monte Carlo simulations using the
Metropolis algorithm on the Kagome lattice with nearest neighbour antiferromagnetic and next-nearest neighbour ferromagnetic interactions. The dynamical exponent $z$ is extracted by fitting to the functional form $cL^z$ at three temperatures at which the system is in the power-law ordered critical phase. Also shown is the anomalous exponent $\eta$ of the power law three-sublattice order at the corresponding temperatures.
}
\end{figure}

\begin{figure}[t]
  \includegraphics[width=8.4cm]{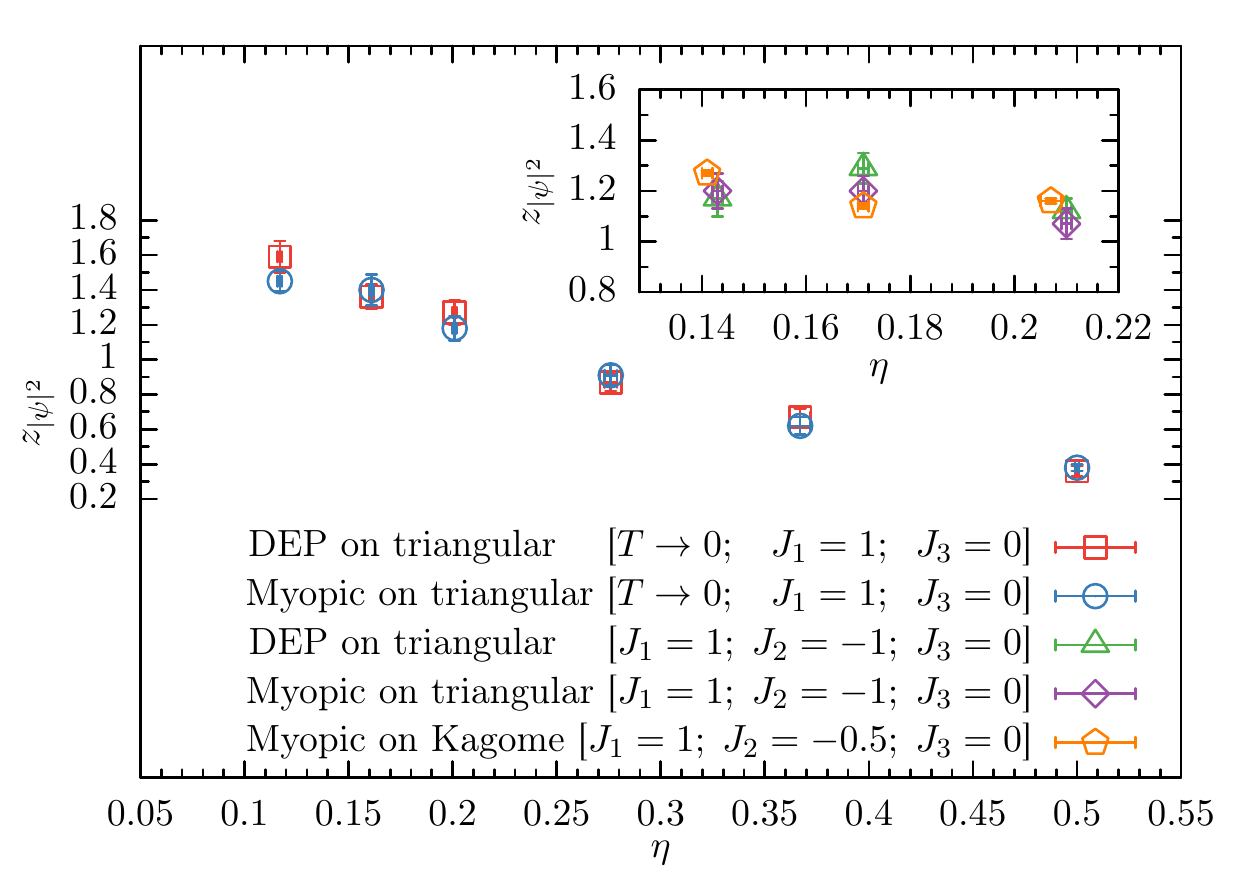}
  \caption{\label{etavsz_psi} The dynamical exponent $z_{|\psi|^2}$ for the three-sublattice
order parameter in simulations employing the DEP and myopic worm algorithms depends in a universal way, independent of the lattice as well as the details of the
worm construction procedure, on the corresponding equilibrium anomalous exponent $\eta$. However, $ z_{|\psi|^2}(\eta)$ appears to be nearly constant in the nonzero
temperature power-law ordered phase, while the corresponding function in the zero
temperature limit is clearly different.}
\end{figure}
\begin{figure}[t]
  \includegraphics[width=8.4cm]{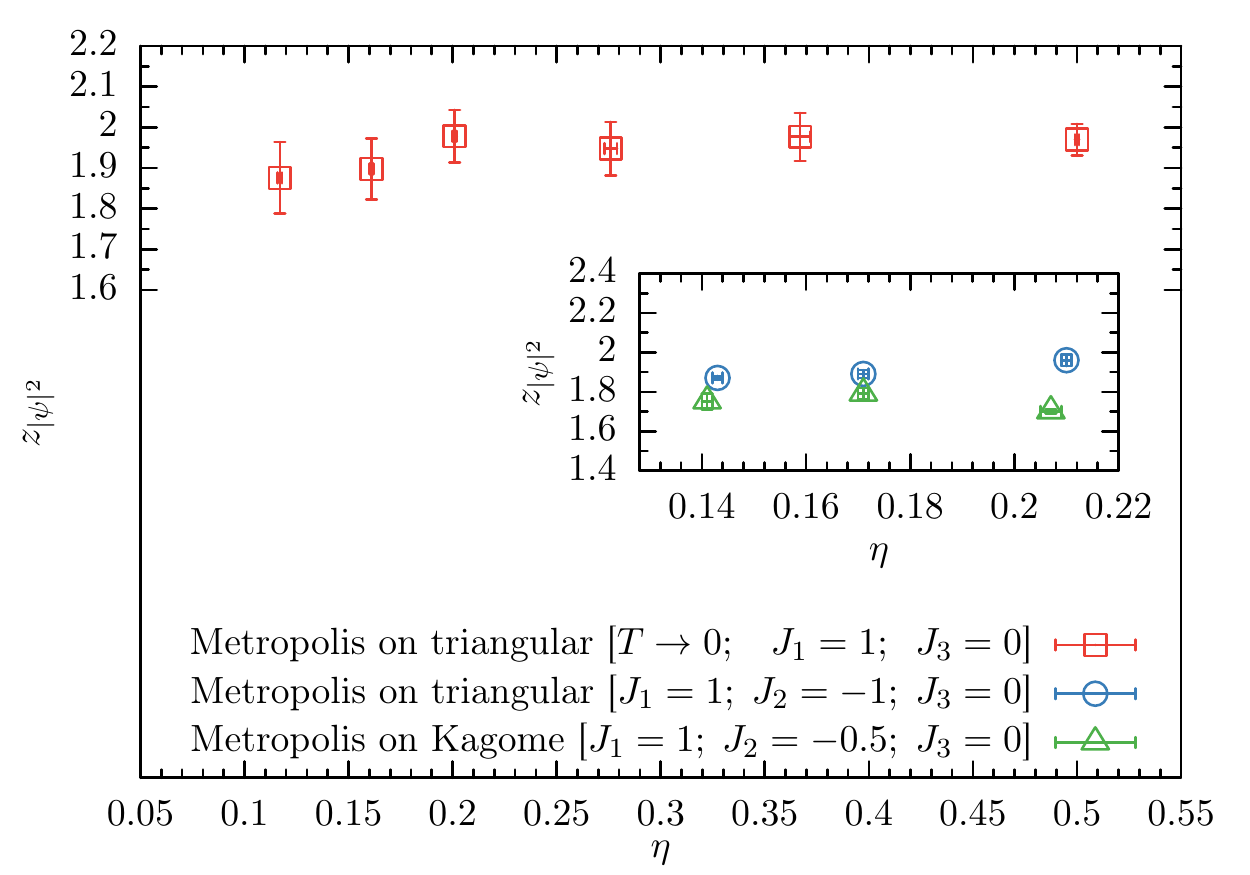}
  \caption{\label{etavsz_psi_metro} Dynamical exponent $z_{|\psi|^2}$ for the three-sublattice
order parameter in simulations employing the Metropolis algorithm.
}
\end{figure}

\section{Performance}
\label{Performance}

As mentioned earlier, the next-nearest and next-next-nearest neighbour interactions induce three-sublattice long-range order of the Ising spins on both triangular \cite{Landau} and Kagome \cite{Wolf_Schotte} lattices. This order melts via a two-step transition with an intermediate power-law ordered critical phase. The correlation function of the three-sublattice order in this phase decays as a power law with a temperature dependent exponent $\eta(T) \in \left( \frac{1}{9},\frac{1}{4} \right)$. Here, we use this power-law ordered intermediate phase associated with two-step
melting of three-sublattice order as a challenging test bed for our algorithms. In
this regime on the triangular lattice, we compare
the performance of the DEP worm algorithm and the myopic worm algorithm with that
of the Metropolis algorithm. Similarly, we compare the performance of
the myopic worm algorithm on the Kagome lattice with that of
the single spin flip Metropolis algorithm in the same critical phase. On the triangular lattice we use $J_1=1$,$J_2=-1$ and $J_3=0$, while for the Kagome lattice system, we use $J_1=1$,$J_2=-0.5$ and $J_3=0$. Monte Carlo simulations are performed on both
lattices for three values of $T$ corresponding to three values of the anomalous exponent $\eta$.

On the triangular lattice in the limiting case of $T \rightarrow 0$ with $J_2=c_2 T$, a power-law phase\cite{Nienhuis_Hilhorst_Blote} is also established for a range of $c_2$. This
$T=0$ phase is characterized by a $c_2$-dependent $\eta(c_2) \in \left( \frac{1}{9},\frac{1}{2} \right) $. With this in mind, we also study this $T \rightarrow 0$ limit on the
triangular lattice, with $J_1=1$ and $J_3=0$, for six values of $c_2 = J_2/T$ corresponding to six values of $\eta$ in this $T=0$ power-law ordered phase.

As is conventional, we characterize the three-sublattice order in terms of
the complex three-sublattice order parameter $\psi \equiv |\psi|e^{i \theta}$. Additionally,
we also monitor  the ferromagnetic order parameter $\sigma$. These
are defined as : $\psi = -\sum_{\vec{R}} e^{i \frac{2\pi}{3} (m+n)} S^z_{\vec{R}}$ and $\sigma= \sum_{\vec{R}}S^z_{\vec{R}}$ on the triangular lattice and $\psi = -\sum_{\vec{R},\alpha} e^{i \frac{2\pi}{3} (m+n-\alpha)} S^z_{\vec{R},\alpha}$ and $\sigma= \sum_{\vec{R},\alpha}S^z_{\vec{R},\alpha}$ on the Kagome lattice, where $\vec{R}=m\hat{e}_x + n\hat{e}_y$ is used to label the sites as shown in Fig.~\ref{lattice} and $\alpha = [0,1,2]$, labels the three basis sites in each unit cell of the Kagome lattice (Fig.~\ref{lattice}). 

In order to meaningfully compare the algorithms, we need a consistent definition of one Monte Carlo step. This is achieved as follows: For the DEP and myopic worm algorithms, we first compute the average number of sites visited by a complete worm of the algorithm at a given point in parameter space. We then adjust the number of worms in one Monte Carlo step (MCS) so that the average number of sites visited in one MCS equals the number of sites in the lattice. For the Metropolis algorithm, we simply fix the number of attempted spin flips in a step to be equal to the number of sites on the lattice, and this defines one MCS for the Metropolis algorithm.

To validate the worm algorithms, we first study the frequency table of the accessed
configurations in long runs on a $3 \times 3$ triangular lattice with the DEP and myopic worm algorithms and a $2 \times 2$ Kagome lattice with the myopic worm algorithm. As is clear from our
results, the measured frequencies are perfectly predicted by theoretical
expectations based on the Boltzmann-Gibbs probabilities of different configurations (See Fig.~\ref{T_freq_allj} for the triangular lattice and Fig.~\ref{K_freq_allj} for the Kagome lattice). 

With all algorithms thus performing an equivalent amount of work in one MCS, we
can compare the performance of these algorithms by measuring the autocorrelation function (defined below) of the order parameter. Given the presence of a critical phase with
power-law three-sublattice order, it is natural to test the performance
of our algorithms by analyzing the $L$ dependence of the autocorrelation time.
More precisely, for a given observable $O$, whose value at the $i^{th}$ MCS is depicted
as $O_i$, we define the normalized autocorrelation function $A_O(k)$ in the standard way: 
\begin{equation}
A_O(k) = \frac{ \langle O_i O_{i+k} \rangle - \langle O_i \rangle \langle O_{i+k}\rangle}{ \langle O_i^2 \rangle - \langle O_i \rangle \langle O_i  \rangle} .
\end{equation}
where $\langle \rangle$ implies averaging over the Monte Carlo run.
We extract autocorrelation times $\tau_{|\psi|^2}$ and $\tau_{\sigma ^ 2}$ by fitting $A(k)$ for the DEP and myopic worm algorithms as well as the Metropolis algorithm at various system sizes $L$ to exponential relaxation functions. We plot these autocorrelation times as a function of $L$ and fit these curves to the power law functional form $cL^z$ in order to extract the corresponding Monte Carlo dynamical exponent $z$. Figs.~\ref{T_gen_ac_psi_j1_j2},~\ref{T_myo_ac_psi_j1_j2},~\ref{T_gen_ac_psi_onlyj2},~\ref{T_myo_ac_psi_onlyj2},~\ref{K_myo_ac_psi_j1_j2} show such power law fits to the $L$ dependence of the autocorrelation time $\tau_{|\psi|^2}$. These fits
provide us our estimates for the dynamical exponent $z_{|\psi|^2}$. Figs.~\ref{T_met_ac_psi_j1_j2},~\ref{T_met_ac_psi_onlyj2} and ~\ref{K_met_ac_psi_j1_j2} show the corresponding plots for the Metropolis algorithm. 

The $\eta$ dependence of the dynamical exponent $z_{|\psi|^2}$ for the DEP and myopic worm algorithms is shown in a comparative plot in Fig.~\ref{etavsz_psi}. 
From these results, it is clear that $z_{|\psi|^2}$ is quasi-universal in the
following sense: In the power-law ordered phase at $T=0$ on the triangular lattice,
$z_{|\psi|^2}$ is independent of the actual details of the worm construction protocol, and
data from the DEP algorithm and the myopic algorithm together define a
$T=0$ functional form $z_{|\psi|^2}(\eta)$. The $T>0$ results are also quasi-universal
in a similar sense: Data from the myopic algorithm on the Kagome and triangular
lattices together define a nearly-constant function $z_{|\psi|^2}(\eta)$. The corresponding
$\eta$ dependence of $z_{|\psi|^2}$ for the Metropolis algorithm is shown in Fig.~\ref{etavsz_psi_metro} in both ($T>0$ and $T=0$) power-law ordered phases. Comparing this
to our results for $z_{|\psi|^2}(\eta)$, we see that both the DEP and myopic worm
algorithms outperform the Metropolis update scheme by a very wide margin.
Corresponding results for the autocorrelation times
of $\sigma^2$ are detailed in the Supplemental Material.

Our $T \rightarrow 0$ results for $z_{|\psi|^2}$ can also be compared to the findings of Ref.~\onlinecite{Zhang_Yang}, in which the efficiency of a plaquette-based generalization
of the Wolff cluster algorithm was studied in the context of the low temperature
limit of the nearest neighbour triangular lattice
Ising antiferromagnet. Comparing to these results, we find that our approach
outperforms this earlier algorithm even in this simple case. Additionally,
this earlier approach does not generalize in an obvious way to systems with
further neighbour couplings, while the approach described here is designed to incorporate such further-neighbour couplings in a straightforward way.

\section{Outlook}
\label{Outlook}
It is natural to wonder if the dual worm strategies introduced here could be ported
to quantum Monte Carlo simulations of frustrated models. Unfortunately, no
straightforward generalization of this type appears possible. The difficulty is
that a worm which alters the state on a given ``time-slice'' of the quantum Monte Carlo
configuration also introduces unphysical off-diagonal terms in the Hamiltonian,
which correspond to ring-exchanges over large loops (in the language of
the dual quantum dimer model). In this context, it is worth nothing that the cluster
construction strategy developed recently for frustrated transverse field Ising models in Ref.~\onlinecite{Biswas_Rakala_Damle} reduces, in the limit of vanishingly small transverse field,
to a version of the Kandel-Ben Av-Domany plaquette-percolation approach.
Thus, generalizations to quantum systems appear to be more natural within that framework, rather than in the dual worm approach used here. Finally,
we note that the quasi-universal
behaviour of the  dynamical exponents $z_{|\psi|^2}$ and $z_{\sigma^2}$ is very
suggestive, throwing up the possibility that the statistics of worms constructed
by these algorithms is universally determined by
the long-wavelength properties of the underlying equilibrium ensemble. We hope to return to this in more detail in a future publication.

\section{Acknowledgements}
We thank D.~Dhar for useful discussions.  Our computational work relied
exclusively on the computational resources of the Department of Theoretical
Physics at the Tata Institute of Fundamental Research (TIFR).

\end{document}